
\input harvmac
\def\rhob{{\rho\kern-0.465em \rho}}

\def\ontopss#1#2#3#4{\raise#4ex \hbox{#1}\mkern-#3mu {#2}}

\setbox\strutbox=\hbox{\vrule height12pt depth5pt width0pt}

\def\strut{\relax\ifmmode\copy\strutbox\else\unhcopy\strutbox\fi}

\nref\rciz{A. Cappelli, C. Itzykson and J.-B. Zuber, Nucl. Phys. B280
(1987) 445.}
\nref\rlm{J. Lepowsky and S. Milne, Adv. in Math. 29 (1978) 271.}
\nref\rfl{A.J. Feingold and J. Lepowsky, Adv. in Math. 29 (1978) 271.}
\nref\rlw{J. Lepowsky and R.L. Wilson, Proc. Nat. Acad. Sci. USA
(1981) 72254.}
\nref\rlp{J. Lepowsky and M. Primc, {\it Structure of the standard
modules for the affine Lie algebra $A_1^{(1)},$}
Contemporary Mathematics, Vol. 46 (AMS, Providence, 1985).}

\nref\rkm{R. Kedem and B.M. McCoy, J. Stat. Phys. 71 (1993),883.}
\nref\rkkmma{R. Kedem, T.R. Klassen, B.M. McCoy and E. Melzer,
Phys. Letts. B304(1993) 263.}
\nref\rkkmmb{R. Kedem, T.R. Klassen, B.M. McCoy and E. Melzer,
Phys. Letts. B 307 (1993) 68.}
\nref\rdkkmm{S. Dasmahapatra, R. Kedem, T.R. Klassen, B.M. McCoy and
E. Melzer, Int. J. Mod. Phys. B (1993) 3617.}
\nref\rdkmm{S. Dasmahapatra, R. Kedem, B.M. McCoy and E. Melzer,
J. Stat. Phys. 74 (1994) 239.}
  \nref\rkmm{R. Kedem, B.M. McCoy, and E. Melzer, The sums of Rogers,
  Schur and Ramanujan and the Bose-Fermi correspondence in $1+1$
  dimensional quantum field theory, hepth 9304056.}

  \nref\rter{M. Terhoeven, Lift of dilogarithm to partition
  identities, BONN-HE-92-36 (1992), hepth 9211120.}
  \nref\rkns{A. Kuniba, T. Nakanishi and J. Suzuki, Mod. Phys. Lett. A8
  (1993) 1835.}

  \nref\rkr{J. Kellendonk and A. Recknagel, Phys. Lett. 298B (1993) 329.}
  \nref\rdas{S. Dasmahapatra, String hypothesis and characters of coset
  CFT's, hepth 9305024.}
  \nref\rfeist{B. Feigin and A. Stoyanovsky, Quasi-particles models for
  the representations of Lie algebras and geometry of flag manifold,
  preprint RIMS 942, hepth 9308079 (1993)}
  \nref\rkrv{J. Kellendonk, M. R{\"o}sgen and R. Varnhagen,
  Int. J. Mod. Phys. A9 (1994) 1009.}

  \nref\rter{M. Terhoeven, Mod. Phys. Lett. A9 (1994) 133.}
  \nref\rmela{E. Melzer, Int. J. Mod. Phys. A9 (1994) 1115.}
  \nref\rmelb{E. Melzer, Lett. Math. Phys. 31 (1994) 233.}
  \nref\rdas{S. Dasmahapatra, Int. J. Mod. Phys. A10 (1995) 875.}
  \nref\rberk{A. Berkovich, Nucl. Phys. B431 (1994) 315.}

  \nref\rfqa{O. Foda and Y-H. Quano, Polynomial Identities of the Rogers
  Ramanujan Type, hepth 9407191.}
  \nref\rfqb{O. Foda and Y-H. Quano, Virasoro character Identities from
  the Andrews-Bailey Construction, hepth 9408086.}
\nref\rfowa{O. Foda and S.O. Warnaar, A bijection which implies
Melzer's polynomial identities: the $\chi_{1,1}^{(p,p+1)}$
case, hepth 9501088.}
\nref\rwa{S.O. Warnaar, Fermionic solution of the Andrews-Baxter-Forrester
model I: Unification of TBA and CTM methods, hepth 9501134.}
  \nref\rwpa{S.O. Warnaar and P.A. Pearce, J. Phys. A27 (1994) L891.}
\nref\rwpb{S.O. Warnaar and P.A. Pearce, A-D-E Polynomial Identities
and the Rogers-Ramanujan Identities, hepth 9411009.}
\nref\rblsa{P. Bouwknegt, A. Ludwig and K. Schoutens, Phys. Lett. B
338 (1994) 488}
\nref\rblsb{P. Bouwknegt, A. Ludwig and K. Schoutens, Spinon basis
for $({\widehat sl_2})_k$ integrable highest weight modules and new
character formulas, to appear in the Proceedings of ``Statistical Mechanics
and Quantum Field Theory,'' USC, May 16-31,1994.}
\nref\rgep{D. Gepner, Lattice models and generalized Rogers-Ramanujan
identities, Phys. Lett. B348 (1995) 377.}
\nref\rbage{E. Baver and D. Gepner, Fermionic sum representations for
the Virasoro characters of the unitary superconformal minimal
models, hepth 9502118.}
\nref\rgeo{G. Georgiev, Combinatorial constructions of modules for
infinite-dimensional dimensional Lie algebras,
I Principal subspace, hepth 9412054; II Parafermionic Space, hepth 9504024.}
\nref\rnya{A. Nakayashiki and Y. Yamada, Crystallizing the spinon basis,
hepth 9504052.}
\nref\rnyb{A. Nakayashiki and Y. Yamada, Crystalline spinon basis for RSOS
models, hepth 9505083.}
\nref\rbm{A. Berkovich and B.M. McCoy, Letts. in Math. Phys. (in press)
hepth 9412030.}
\nref\rmel{E. Melzer, Supersymmetric Analogs of the Gordon-Andrews
Identities and Related TBA Systems, hepth 9412154}

\nref\rbosform{B.L. Feigin and D.B. Fuchs, Funct. Anal. Appl. 16 (1982) 114;
P. Goddard, A. Kent and D. Olive, Comm. Math. Phys. 103 (1986) 105;
A. Meurman and A. Rocha-Caridi, Comm. Math. Phys. 107 (1986) 263.}
\nref\rand{G.E. Andrews, in {\it The theory and applications of
special functions}, ed. R. Askey (Academic Press, New York, 1975)}
\nref\rbres{D.M. Bressoud, Quart. J. Math. Oxford (2) 31 (1980) 385.}
\nref\rslat{L. Slater, Proc. Lond. Math. Soc. (2) 54 (1952) 147.}
\nref\rgol{H. G{\"o}llnitz, J. Reine Angew. Math. 225 (1967) 154.}
\nref\rgor{B. Gordon, Duke. Math. J. 31 (1965) 741.}

\nref\rburge{W.H. Burge, Europ. J. Combinatorics 3 (1982) 195.}
\nref\rab{G.E. Andrews and R.J. Baxter, J. Stat. Phys. 47 (1987) 297.}
\nref\randb{G.E. Andrews, J. Am. Math. Soc. 3 (1990) 653.}
\nref\randc{G.E. Andrews, Contemporary Mathematics 166 (1994) 141.}
\nref\rwit{E. Witten, Nucl. Phys. B202 (1982) 253.}
\nref\rehol{W. Eholzer and R. H{\" u}bel, Nucl. Phys. B414 (1994) 348.}
\nref\rrog{L.J. Rogers, Proc. London Math. Soc. (2), 16 (1917) 315.}
\nref\randd{G.E. Andrews, {\it Partitions: yesterday and today}, New
Zealand Math. Soc., Wellington, 1979.}

\nref\rts{M. Takahashi and M. Suzuki, Prog. of Theo. Phys. 48 (1972) 2187.}
\nref\rdsz{P. DiFrancesco, H. Saleur and J.-B. Zuber, Nucl. Phys. 300
(1988) 393.}
\nref\rande{G.E. Andrews, {\it Theory of Partitions}, Addison-Wesley,
Reading Mass. (1976).}
\nref\rrc{A. Rocha-Caridi,in {\it Vertex Operators in mathematics and physics,}
ed. J. Lepowsky, S. Mandelstam and I.M. Singer,(Springer, Berlin, 1985).}
\nref\rfb{P.J. Forrester and R.J. Baxter, J. Stat. Phys. 38 (1985) 435.}
\nref\dmbres{D.M. Bressoud, in {\it Ramanujan Revisited}, pp. 57-67,
G.E. Andrews et al eds., Academic Press, 1988.}
\nref\rfatzam{V.A. Fateev and Al.B. Zamolodchikov, Phys. Lett. B271 (1991) 91.}
\nref\rabf{G.E. Andrews, R.J. Baxter and P.J. Forrester, J. Stat. Phys. 35
(1984) 193.}
\nref\rtat{R. Tateo, New functional dilogarithm identities and sine-Gordon
Y-systems, hepth 9505022.}
\nref\rrst{F. Ravanini, M. Staniskov and R. Tateo, Integrable perturbations of
CFT with complex parameter: the $M_{3/5}$ model and its generalizations,
hepth 9411085.}
\nref\rfeino{B.L. Feigin, T. Nakanishi and H. Ooguri, Int. J. Mod. Phys. A7,
Suppl. 1A (1992) 217.}
\nref\andf{G.E. Andrews, Rogers-Ramanujan polynomials for modulus 6,
(preprint).}

  \Title{\vbox{\baselineskip12pt\hbox{BONN-TH-95-12}
  \hbox{ITPSB 95-18}
  \hbox{HEP-TH 9507072}}}
  {\vbox{\centerline{Polynomial Identities, Indices, and Duality}
  \centerline{for the}
  \centerline{$N=1$ Superconformal Model $SM(2,4\nu)$}}}
  \centerline{Alexander Berkovich~\foot{berkovic@axpib.physik.uni-bonn.de}}

  \bigskip\centerline{\it Physikalisches Institut der}
  \centerline{\it Rheinischen Friedrich-Wilhelms Universit{\"a}t Bonn}
  \centerline{\it Nussallee 12}
  \centerline{\it D-53115 Bonn, Germany}
  \bigskip
  \centerline{and}
  \bigskip
  \centerline{ Barry~M.~McCoy~\foot{mccoy@max.physics.sunysb.edu} and
               William~P.~Orrick\foot{worrick@insti.physics.sunysb.edu}}

  \bigskip\centerline{\it Institute for Theoretical Physics}
  \centerline{\it State University of New York}
  \centerline{\it Stony Brook,  NY 11794-3840}
  \bigskip
  \centerline{\it Dedicated to the memory of Claude Itzykson}
  \Date{\hfill 7/95}

  \eject

\centerline{\bf Abstract}

We prove polynomial identities for the $N=1$
superconformal model $SM(2,4\nu)$ which generalize and extend the
known Fermi/Bose character identities.  Our proof uses the $q$-
trinomial coefficients of Andrews and Baxter on the bosonic side and
a recently introduced very general method of producing recursion
relations for $q$-series on the fermionic side. We use these
polynomials to demonstrate a dual relation under $q\rightarrow q^{-1}$
between $SM(2,4\nu)$ and $M(2\nu-1,4\nu)$. We also introduce a generalization
of the Witten index which is expressible in terms of the Rogers false theta
functions.

\newsec{Introduction}

All chiral partition functions of conformal field theory have two
distinct representations; 1) a bosonic form which may be expressed
in terms of theta functions from which modular transformation
properties are readily apparent~\rciz~and 2) a fermionic form in terms of
$q$-series in which the quasiparticle spectrum of the theory is
clearly seen. The bosonic form is most useful in computing the
conformal dimensions. The fermionic form is best adapted
to study massive perturbations. The equality of the two forms can be
thought of as generalized Rogers-Ramanujan identities.

The study of the bosonic representations has been well developed for
over a decade. However, with the exceptions of the pioneering work on
characters of $A^{(1)}_1$~\rlm--\rfl~and the $Z_N$ parafermionic
theories~\rlw--\rlp~the study of the fermionic representations started only
several years ago and in the last few years there have been many
conjectures and proofs of fermionic representations of the various characters
{}~\rkm--\rmel.

In this paper we consider the $N=1$ superconformal model $SM(2,4\nu).$
The bosonic form of this model's characters is a special case of the general
formula~\rbosform
\eqn\schi{{\hat\chi}^{(p,p')}_{r,s}(q)={\hat\chi}_{p-r,p'-s}^{(p,p')}(q)=
{(-q^{\epsilon_{r-s}})_{\infty}\over
(q)_{\infty}}\sum_{j=-\infty}^{\infty}\bigl( q^{{j(jpp'+rp'-sp)\over 2}}
-q^{{(jp+r)(jp'+s)\over 2}}\bigr)}
where
\eqn\qden{(A)_k=\cases{\prod_{j=0}^{k-1}(1-A q^j),& $k=1,2,\cdots$ \cr
1,& k=0 \cr}}
and
\eqn\epsden{\epsilon_a=\cases{{1\over2}&if $a$ is even (Neveu-Schwarz
(NS) sector)\cr 1& if $a$ is odd (Ramond (R) sector)\cr}}
Here $r=1,2,\cdots,p-1$ and $s=1,2,\cdots,p'-1$ and $p$ and ${(p'-p)\over 2}$
are coprime.

Setting $p=2$, $p'=4\nu$ in~\schi~we have for $n=0,\pm 1$
\eqn\bdef{{\hat\chi}^{(2,4\nu)}_{1,2\nu-2s'+|n|-1}(q)
\equiv B^{(\nu,n)}_{s'}(q)={(-q^{{1+|n|\over 2}})_{\infty}\over
(q)_{\infty}}\sum_{j=-\infty}^{\infty}(-1)^j q^{\nu j^2+j(s'+{1-|n|\over 2})}}
where here and throughout the rest of the paper
\eqn\srange{s'=0,1,2,...,\nu-1}
and $n=0(\pm 1)$ corresponds to the $NS(R)$ sector.

The Fermionic representations of $SM(2,4\nu)$ characters are given in terms
of the function $F^{(\nu,n)}_{s'}(q)$ defined for $n=0,\pm 1$ as follows
\eqn\fdefn{F^{(\nu,n)}_{s'}(q)=\sum_{m_1,n_2,\cdots,n_{\nu}\geq 0}
{q^{Qf+Lf_{n,s'}}\over(q)_{n_2}(q)_{n_3}\cdots (q)_{n_\nu}}{N_2
\atopwithdelims[]m_1}_q,}
where
the $q$-binomial coefficient is defined in a slightly unconventional way as
\eqn\qbin{{l\atopwithdelims[] m}_q=\cases{{(q)_l\over (q)_m
(q)_{l-m}}&if $0\leq m \leq l$\cr
1&if $m=0,~l\leq -1$\cr
0& otherwise,\cr}}
the quadratic form $Qf$ and linear form $Lf_{n,s'}$ are
\eqn\qf{Qf={m_1^2\over 2}-m_1 N_2+\sum_{j=2}^{\nu} N^2_j}

\eqn\lf{Lf_{n,s'}=n{m_1\over 2}+\sum_{l=\nu-s'+1}^{\nu} N_l,}
with
\eqn\Ndef{N_k=\sum_{j=k}^{\nu}n_j.}

Once again $n=0$ corresponds to the $NS$ sector and $n=+1(-1)$ corresponds to
the first (second) representation for the Ramond sector which we will call
$R^+(R^-)$. We note in passing that the reason for existence of these two
representations can be traced back to the fact that zero-modes of fermionic
fields act nontrivially on the highest weight vectors.

The relation between the bosonic and fermionic forms depends on the
characters studied.  We consider three separate cases

1) For the Neveu-Schwarz sector we have:
\eqn\bfns{B_{s'}^{(\nu,0)}(q)=F_{s'}^{(\nu,0)}(q);}

2) For $R^{+}$
\eqn\bfrp{B_{s'}^{(\nu,1)}(q)=\cases{{1\over 2}(F^{(\nu,1)}_{s'}(q)+
F^{(\nu,1)}_{s'-1}(q))& for $ s'\neq 0 $\cr
F^{(\nu,1)}_{0}(q)& for $s'=0;$\cr}}

3) For $R^{-}$
\eqn\bfrm{F_{s'}^{(\nu,-1)}(q)=\cases{B_{\nu-1}^{(\nu,-1)}(q)&for
$s'=\nu-1$\cr
B_{s'}^{(\nu,-1)}(q)+B_{s'+1}^{(\nu,-1)}(q)&for $s'\neq\nu-1$\cr}}
or, equivalently
\eqn\bfrmt{B_{s'}^{(\nu,-1)}(q)=\sum_{l=s'}^{\nu-1}
(-1)^{l+s'}F_{l}^{(\nu,-1)}(q).}

In the Neveu-Schwarz sector the identities~\bfns~are the generalizations
to arbitrary $\nu$ by Andrews~\rand~(for $s'=0$) and Bressoud~\rbres~of
the $\nu=2$ results due to Slater (eqns. (34), (36) of~\rslat)
also known as G{\"o}llnitz-Gordon identities~\rgol~and~\rgor.
For the Ramond cases $R^-$ with $s'=\nu-1$ and $R^+$
with $s'=0$ the identities~\bfrp~and~\bfrm~have
been conjectured by Melzer~\rmel. For all other values of $s'$ the
results of~\bfrp~and~\bfrm~are new. In general, the Ramond sector Fermi forms
should also be compared with the result of Burge~\rburge~(stated at the
bottom of page 204 with the misprint $(q^2,q^2)_{n_{k-1}}$
corrected to $(q^2,q^2)_{n_{k-2}}$) where a free Fermi term is
factored out and the number of variables in the sum is reduced to $\nu-1.$
A direct proof of the equivalence of~\bfrp~with~\rburge~does not
seem to be known.

The first purpose of this paper is to generalize both the bosonic and the
fermionic  expressions from infinite series to polynomials. Indeed, we
will see that there are not one, but many distinct polynomials
which generalize~\schi~and~\fdefn. We will then prove Fermi/Bose identities
for these polynomials by obtaining recursion relations between
several different polynomials which are related to a given character.
These polynomial identities will reduce to~\bfns--\bfrm~when the degree of
polynomials goes to infinity.
Our tools in this proof will be the use of the $q$-trinomial coefficients
of Andrews and Baxter~\rab--\randc~on the bosonic side and the methods of
ref.~\rberk~ on the fermionic side.

By the very name the $N=1$ superconformal models have an
interpretation in terms of a fermion and a boson, and one aspect of
this interpretation is seen in the factorization of the bosonic
form~\schi~into a free fermionic factor $(-q^{\epsilon_{r-s}})_\infty$
and another factor which looks as if it is obtained from a free boson by
projecting out null states. Correspondingly, there should be an
interpretation of the Fermi form~\fdefn~which separates the
quasiparticles into one which represents the fermion and the rest
which represent what in the bosonic form was called the projected
boson. One such interpretation is instantly suggested by the
form~\fdefn~itself where $m_1$ and $n_i$ appear in quite different
ways. We will thus adopt the tentative interpretation that $m_1$ is
related to the fermion number operator $F$ or perhaps more accurately that
$(-1)^{m_1}$ is related to the chirality operator $(-1)^F.$
With this identification we can consider the object
\eqn\index{{\tilde F}^{(\nu,n)}_{s'}(q)=
\sum_{m_1,n_2,\cdots,n_{\nu}\geq 0}{(-1)^{m_1}q^{Qf+Lf_{n,s'}}\over
(q)_{n_2}(q)_{n_3}\cdots(q)_{n_{\nu}}}{N_2\atopwithdelims[]
m_1}_q}
and ask what relation it has with
\eqn\windex {{\rm Tr}(-1)^F \exp(-H).}

In the $NS$ sector this relation is straightforward. Replacing $\sqrt q$
by $-\sqrt q$ in~\bfns~we immediately note
\eqn\nssector{{\tilde F}_{s'}^{(\nu,0)}(q)=F_{s'}^{(\nu,0)}(e^{2\pi i}q)
=B_{s'}^{(\nu,0)}(e^{2\pi i}q)}
Clearly, ${\tilde F}_{s'}^{(\nu,0)}(q)$ is the $T$-modular transform of
$F_{s'}^{(\nu,0)}(q)$ and therefore must be equal to~\windex~according to
{}~\rehol.
In the Ramond  sector we again find that there are two distinct
cases. For $R^+$ we define in analogy with~\bfrp~
\eqn\tbfrp{{\tilde B}_{s'}^{(\nu,1)}(q)=\cases{{1\over 2}
({\tilde F}^{(\nu,1)}_{s'}(q)-
{\tilde F}^{(\nu,1)}_{s'-1}(q))& for $ s'\neq 0 $\cr
{\tilde F}^{(\nu,1)}_{0}(q)& for $s'= 0.$\cr}}
Then since we prove in sec.~5 that
\eqn\exramsec{{\tilde F}_{s'}^{(\nu,1)}(q)=1}
we see that
\eqn\tbwit{{\tilde B}_{s'}^{(\nu,1)}(q)=
\cases{0&for $s'\neq 0$\cr
1&for $s'=0$\cr}}
which is equal to the Witten indices~\rwit~as studied in~\rehol.
We want to emphasize that formulas~\tbfrp~are not identities, but definitions.
However, in sec.~5 we will find polynomial identities  for $s'\neq 0$, which
provide extra motivation for the definitions above. For the case
$s'=0$ an  appropriate polynomial identity is still lacking.
Our  motivation in this
case is the analogy with~\bfrp~and the fact that we have an agreement with the
Witten index calculations of~\rehol.

For the Ramond case of $R^-$ we  define in analogy with~\bfrmt
\eqn\tbfrmt{{\tilde B}_{s'}^{(\nu,-1)}(q)=\sum_{l=s'}^{\nu-1}
{\tilde F}_{l}^{(\nu,-1)}(q).}
In sec.~5 we find the bosonic companion of ${\tilde F}^{(\nu,-1)}_{s'}(q)$.
Remarkably, it is not a constant, but rather is
\eqn\tfnlim{{\tilde F}_{s'}^{(\nu,-1)}(q)=\cases{I_{\nu-1}^{(\nu)}(q)&for
$s'=\nu-1$\cr
I^{(\nu)}_{s'}(q)
-I^{(\nu)}_{s'+1}(q)&for $s'\neq \nu-1$\cr}}
where
\eqn\false{I_{s'}^{(\nu)}(q)=1+\sum_{j=1}^{\infty}q^{{\nu}j^2}(q^{s'j}-
q^{-s'j})}
is the false theta function introduced by Rogers~\rrog~and extensively
studied by Andrews~\randd.
Thus
\eqn\tblim{{\tilde B}_{s'}^{(\nu,-1)}(q)=I_{s'}^{(\nu)}(q).}
We show in sec.~5  that
\eqn\limfal{\lim_{q \rightarrow 1}I_{s'}^{(\nu)}(q)=1-{s'\over\nu}}
which suggests that it is possible to define for $R^-$~a
fractional analogue of the Witten index.

In sec.~2 we will state in detail the polynomial analogs of
identities~\bfns--\bfrm~and the sets of recursion relations we will use to
prove them.  In sec.~3 we will show that the fermionic polynomials satisfy
these recursion relations and in sec.~4 we will do it for the bosonic
polynomials. In sec.~5 we will discuss the Fermi forms
${\tilde F}_{s'}^{(\nu,\pm 1)}(q)$ and the indices
${\tilde B}^{(\nu,\pm 1)}_{s'}(q).$ In sec.~6 we will use the polynomial
identities to study the dual relation which exists between
$SM(2,4\nu)$ and $M(2\nu-1,4\nu)$ under the replacement $q\rightarrow q^{-1}.$
Finally in sec.~7 we will discuss representation theoretical consequences of
two partition identities due to Burge. We will conclude with some remarks about
possible generalizations and open questions. Technical details concerning
$q$-trinomial coefficients will be treated in the appendix for continuity of
presentation.

\newsec{Polynomials and Recursion relations.}

The starting point for proving Rogers-Ramanujan type identities by the
method of ~\rberk~ is identifying an $({\vec n},{\vec m})$-system
and an associated counting problem. For the present case the appropriate
$({\vec n},{\vec m})$-system is as follows
\eqn\countg{\eqalign{n_1+m_1&={1\over 2}(L+m_1-m_2)-a_1\cr
n_2+m_2&={1\over 2}(L+m_1+m_3)-a_2\cr
n_i+m_i&={1\over 2}(m_{i-1}+m_{i+1})-a_i,~~~~{\rm for}~3\leq i \leq \nu-1\cr
n_{\nu}+m_{\nu}&={1\over 2}(m_{\nu-1}+m_{\nu})-a_{\nu}\cr}}
where $n_i$ and $m_i$ are integers and the components $a_i$  of the vector
${\vec a}$ are either integers or half integers.
This system is closely related to the $TBA$ equations for the $XXZ$-model
((3.9) of~\rts) with anisotropy
\eqn\anis{\gamma=\pi{(2\nu-1)\over 4\nu}.}
In the language of our previous treatment ~\rbm~ of the $M(p,p')$ minimal
models, system~\countg~consists of two Takahashi zones with tadpoles at the
end of each zone. The principal difference between the present case and
the one considered in ~\rbm~ is the appearance of two inhomogeneous terms
${L\over 2}$ in the first and second equations. The second inhomogeneous term
arises because $N=1$ superconformal models are derived from the spin $1$
$XXZ$ chain~\rdsz~while $N=0$ models, investigated in~\rbm,~are derived
from the spin ${1\over 2}$ chain. The presence of the first term in~\countg~
indicates that the spin $1$ $XXZ$ model with $\gamma$ given by~\anis~is in the
regime of strong anisotropy. This inhomogeneous term is not expected to
be present for any other $N=1$ $SM(p,p')$ model with ${2p'\over p'-p}\geq 3$.

The $({\vec n},{\vec m})$-system~\countg~describes $\nu$ Fermi bands.
Each band consists of $n_i+m_i$ consecutive integers with only $n_i$
distinct integers being occupied by the $n_i$ quasiparticles. The remaining
$m_i$ integers can be thought of as holes. If one allows particles to move
freely in each band (subject only to fermionic exclusion rules) then one is
naturally led to the following counting problem
\eqn\fcount{F(L)=\sum_{n_i,m_i\geq 0}\prod_{i=1}^{\nu}
{n_i+m_i\atopwithdelims[] n_i}}
where the summation variables $n_i,m_i$ are related by~\countg~ with ${\vec a}$
fixed to be zero for the time being. To calculate $F(L)$ we use three
simple consequences of~\countg
\eqn\fcons{L=n_1+m_{\nu}+\sum_{i=2}^{\nu}(2i-3)n_i}

\eqn\scons{m_i=m_{\nu}+2\sum_{j=i+1}^{\nu}N_j,~~~~~i\geq 2}

\eqn\tcons{n_1+m_2=N_2}
along with the generating function technique (sec.~2 of ~\rberk) to obtain
\eqn\clfbeqn{F(L)=B(L)}
where
\eqn\clbosform{B(L)=\sum_{j=-\infty}^{\infty}(-1)^j
\biggl({L\atopwithdelims() 2\nu j}_2+{L\atopwithdelims() 2\nu j+1}_2\biggr)}
and $N_j$ was defined in~\Ndef.

The trinomial coefficients ${L\atopwithdelims() i}_2$ which appear in the
above equation are given by
\eqn\cltrdef{(z+1+{1\over z})^L=\sum_{i=-L}^L{L\atopwithdelims()i}_2 z^i.}

In what follows we will consider three different $q$-analogs of~\clfbeqn~
associated with the $NS$ and $R^{\pm}$ sectors. We remark that these
$q$-deformations amount to prescribing the linear dispersion law for the
quasiparticles described above. We also point out that one can use~\countg~and
{}~\fcons~to find a pictorial representation for quasiparticles in the spirit
of~\rwa. This representation will be given elsewhere.

Motivated by~\fcount~ we now introduce the polynomial generalization of
the fermionic form $F_{s'}^{(\nu,n)}(q)$~\fdefn-\lf
\eqn\fpoly{F^{(\nu,n)}_{r',s'}(L,q)=\sum_{{\cal D}_{r',s'}} q^{Qf+Lf_{n,s'}}
\prod_{i=1}^{\nu}{n_i+m_i\atopwithdelims[] n_i}_q}
where the ``finitization'' parameter $r'$ is
\eqn\rrange{r'=0,1,2,\cdots,\nu-2}
and the variables $n_i,m_i$ are related by~\countg~with the vector ${\vec a}$
defined by
\eqn\indefn{\eqalign{{\vec a}&={\vec a}^{(r')}+{\vec a}^{(s')}\cr
a^{(k)}_i&=\cases{{1\over 2}
(\delta_{i,{\nu}}-\delta_{i,\nu-k})&for $0\leq k \leq \nu-2$\cr
{1\over 2}(\delta_{i,\nu}+\delta_{i,1}) &for $k=\nu-1$.\cr}}}

The domain of summation, ${\cal D}_{r',s'}$ is best described in terms of
${\vec n}$ and $m_{\nu}$ which are subject to the constraint
\eqn\constraint{L=(n_1+a_1)+m_{\nu}+\sum_{i=2}^{\nu}(2i-3)(n_i+a_i).}
All other variables are given by
\eqn\eqnva{m_1=N_2-n_1}
\eqn\eqnvb{m_i=m_{\nu}+2\sum_{j=i+1}^{\nu} (j-i)(n_j+a_j),
{}~~~~~i=2,3,\cdots,\nu-1.}
Keeping in mind that
\eqn\qbinnewdef{{{\rm neg.~ int.}\atopwithdelims[] 0}_q=1,}
we define ${\cal D}_{r',s'}$ for $s'\geq r'$ as the union of the sets
of solutions to~\constraint~satisfying
\eqn\extra{\eqalign{0:~~~&n_i,~m_{\nu}\geq 0,\cr
1:~~~&n_{\nu}=0,m_{\nu}=-2,~n_1,\cdots, n_{\nu-1}\geq 0,\cr
2:~~~&n_{\nu}=n_{\nu-1}=0,~m_{\nu}=-4,~n_1,\cdots n_{\nu-2}\geq 0,\cr
&~~~~\cdots\cr
r':~~~&n_{\nu}=n_{\nu-1}=\cdots =
n_{\nu-r'+1}=0,~~~m_{\nu}=-2r',~~~n_1,\cdots,n_{\nu-r'}\geq 0;\cr}}
and for $s'<r'$ the definition is the same  as above with $r'\rightarrow s'$.

Using the asymptotic formula
\eqn\qbinlim{\lim_{A\rightarrow\infty}{A\atopwithdelims[]B}_q={1\over(q)_B}}
and the simple consequence of~\countg
\eqn\ancons{n_i+m_i=L+m_1+n_i-2\sum_{j=2}^i (j-1)(n_j+a_j)-
2\sum_{j=i+1}^{\nu} (i-1)(n_j+a_j);~~~i\geq 2}
along with~\eqnva, we establish relations between $F_{r',s'}^{(\nu,n)}(L,q)$
and the fermionic forms~\fdefn
\eqn\pcf{\lim_{L\rightarrow\infty}F_{r',s'}^{(\nu,n)}(L,q)=F^{(\nu,n)}_{s'}(q)}
which hold for all $r'.$

To write the bosonic polynomials one needs the $q$-analogs of the trinomial
coefficients ${L\atopwithdelims()A}_2$ introduced in~\cltrdef.
Following Andrews and Baxter~\rab~we define
\eqn\appa{{L,A-n;q\atopwithdelims()A}_2=\sum_{j \geq 0}t_n(L,A;j),~~~~~n\in Z}
and
\eqn\tndf{T_n(L,A;q^{{1\over 2}})=q^{{L(L-n)-A(A-n)\over 2}}
{L,A-n;q^{-1}\atopwithdelims() A}_2}
where
\eqn\appaa{t_n(L,A;j)={q^{j(j+A-n)}(q)_L\over (q)_j (q)_{j+A}(q)_{L-2j-A}}.}
We note the elementary property
\eqn\prop{T_n(L,A;q^{{1\over 2}})=T_n(L,-A;q^{{1\over 2}})}
and remark that
\eqn\tnsym{T_n(L,A;-q^{{1\over 2}})=\cases{(-1)^{L+A}T_n(L,A;q^{{1\over 2}})
&for $n$ even\cr
T_n(L,A;q^{{1\over 2}})& for $n$ odd\cr}}
Consequently, $T_n(L,A;q^{{1\over 2}})$ is actually a polynomial in $q$ for
$n$ odd or for $n$ even and $L+A$ even, while for $n$ even
and $L+A$ odd $T_n(L,A,q^{{1\over 2}})$ contains only odd powers of
$q^{{1\over 2}}.$

We then have the following definition of bosonic polynomials:

1) For the Neveu-Schwarz sector
\eqn\bosepolyns{\eqalign{B^{(\nu,0)}_{r',s'}(L,q)=&
\sum_{j=-\infty}^{\infty}(-1)^j q^{\nu
j^2+(s'+{1\over 2})j}\biggl(T_0(L,2\nu j +s'-r';q^{{1\over 2}})\cr
&+T_0(L,2\nu j +s'+1+r';q^{{1\over 2}})\biggr);}}

2) For the  Ramond sector $R^+$
\eqn\bosepolyrp{\eqalign{&B^{(\nu,1)}_{r',s'}(L,q)=\cr
&{1\over 2}\sum_{j=-\infty}^{\infty}(-1)^j q^{\nu
j^2+s'j}\biggl(T_{-1}(L,2\nu j +s'-r';q^{{1\over 2}})+
T_{-1}(L,2\nu j +s'+1+r';q^{{1\over 2}})\cr
&~~~~~~~~~~+T_{-1}(L,2\nu j +s'-r'-1;q^{{1\over 2}})+\
T_{-1}(L,2\nu j +s'+r';q^{{1\over 2}})\biggr);\cr}}

3) For the  Ramond sector $R^-$
\eqn\bosepolyr{B^{(\nu,-1)}_{r',s'}(L,q)=\sum_{j=-\infty}^{\infty}
(-1)^j q^{\nu j^2+s'j}\sum_{i=-r'}^{r'}
(-1)^{r'+i}T_1(L,2\nu j+s'+ i;q^{{1\over 2}})}
where $s'=0,1,2,\cdots,\nu-1$ and $r'=0,1,2,\cdots,\nu-2$.

Using the limiting formula of the appendix
\eqn\tzlim{\lim_{L\rightarrow \infty}T_n(L,A;
q^{{1\over 2}})=\cases{{(-q^{{(1-n)\over 2}})_{\infty}+
(q^{{(1-n)\over 2}})_{\infty}\over 2(q)_{\infty}}&if $L-A$ is even\cr
{(-q^{{(1-n)\over 2}})_{\infty}-
(q^{{(1-n)\over 2}})_{\infty}\over 2(q)_{\infty}}& if $L-A$ is odd.\cr}}
and noting the special case
\eqn\tolim{\lim_{L\rightarrow \infty}
T_{1}(L,A;q^{{1\over 2}})={(-q)_{\infty}\over (q)_{\infty}}}
we find the relation between the polynomials $B^{(\nu,n)}_{r',s'}(L,q)$
and the characters~\bdef
\eqn\pcb{\eqalign{\lim_{L\rightarrow\infty}B^{(\nu,0)}_{r's'}(L,q)&=
B^{(\nu,0)}_{s'}(q)\cr
\lim_{L\rightarrow\infty}B_{r',s'}^{(\nu,1)}(L,q)&=
\lim_{L\rightarrow\infty}B_{r',s'}^{(\nu,-1)}(L,q)
=B^{(\nu,\pm 1)}_{s'}(q)\cr}}
which holds for all $r'.$

We will prove the following polynomial identities which generalize
the character identities ~\bfns--\bfrm

1) For $NS$
\eqn\polyidenta{F^{(\nu,0)}_{r',s'}(L,q)=B^{(\nu,0)}_{r',s'}(L,q);}

2) For $R^+$
\eqn\polyidentb{B^{(\nu,1)}_{r',s'}(L,q)=\cases{{1\over 2}(F^{(\nu,1)}_{r',s'}
(L,q)+F^{(\nu,1)}_{r',s'-1}(L,q))&for $s'\neq 0$\cr
F_{r',0}^{(\nu,1)}(L,q)&for $s'=0;$\cr}}

3) For $R^-$
\eqn\polyidentc{F^{(\nu,-1)}_{r',s'}(L,q)=\cases{B^{(\nu,-1)}_{r',\nu-1}(L,q)
&for $s'=\nu-1$\cr
B^{(\nu,-1)}_{r',s'}(L,q)+B^{(\nu,-1)}_{r',s'+1}(L,q)
&for $s'\neq\nu-1.$\cr}}
by showing that both $F^{(\nu,n)}_{r',s'}(L,q)$ and $B^{(\nu,n)}_{r',s'}(L,q)$
satisfy the following set of recursion relations for $\nu\geq 3$ in the
variables $L$ and $r'$
\eqn\nseqn{\eqalign{h_0(L)&=
h_1(L-1)+(q^{L-{1-n\over2}}+1)h_0(L-1)+(q^{L-1}-1)h_0(L-2),\cr
h_r(L)&=h_{r-1}(L-1)+h_{r+1}(L-1)+q^{L-{1-n\over 2}}h_r(L-1)+
(q^{L-1}-1)h_r(L-2)\cr
&~~~~~~~~~~~~~~~~~~~~~~~~~~~~~~~~~~~~~~~~~~~~~~~{\rm for}~1\leq r\leq \nu-3,\cr
h_{\nu-2}(L)&=h_{\nu-3}(L-1)+q^{L-{1-n\over 2}}h_{\nu-2}(L-1)+
q^{L-1}h_{\nu-2}(L-2);\cr}}
where
\eqn\ndefn{n=\cases{~~0& for $NS$\cr
{}~~1& for $R^+$\cr
-1&for $R^-.$\cr}}
Note that the first and the last equations follow from the middle equation
if one introduces $h_{-1}(L)$ and $h_{\nu-1}(L)$ satisfying
\eqn\hmo{h_{-1}(L)=h_{0}(L)}
and
\eqn\hnmo{h_{\nu-1}(L)=h_{\nu-2}(L-1).}
For $\nu=2$ there is only the single equation
\eqn\nutns{h_0(L)=(1+q^{L-{1-n\over 2}})h_0(L-1)+q^{L-1}h_0(L-2).}

Observe that the recursion relations in the sectors $NS$ and $R^{\pm}$
are independent of $s'$. The proof of the polynomial identities will be
completed by showing that~\polyidenta--\polyidentc~hold for $L=0,1$.
We record here the values of the fermionic and bosonic forms at
$L=0,1$, computed directly from~\fpoly~and~\bosepolyns--\bosepolyr.
Notice that there is no dependence on $\nu$.  The fermionic
forms are
\eqn\fbc{\eqalign{F^{(\nu,n)}_{r',s'}(0,q)=&\delta_{r',s'},\cr
F^{(\nu,0)}_{r',s'}(1,q)=&\cases{1+q^{1\over 2} &if $r'=s'=0$\cr
q^{1\over 2} &if $r'=s'\ge 1$\cr
1 &if $r'=s'+1$ or $s'=r'+1$ \cr
0 &otherwise,\cr}\cr
F^{(\nu,1)}_{r',s'}(1,q)=&\cases{1+q &if $r'=s'=0$\cr
q &if $r'=s'\ge 1$\cr
1 &if $r'=s'+1$ or $s'=r'+1$ \cr
0 &otherwise,\cr}\cr
F^{(\nu,-1)}_{r',s'}(1,q)=&\cases{2 &if $r'=s'=0$\cr
1 &if $r'=s'\ge 1$ or $r'=s'+1$ or $s'=r'+1$ \cr
0 &otherwise.\cr}\cr}}
The bosonic forms are
\eqn\bbc{\eqalign{B^{(\nu,0)}_{r',s'}(0,q)=&\delta_{r',s'},\cr
B^{(\nu,0)}_{r',s'}(1,q)=&\cases{1+q^{1\over 2} &if $r'=s'=0$\cr
q^{1\over 2} &if $r'=s'\ge 1$\cr
1 &if $r'=s'+1$ or $s'=r'+1$ \cr
0 &otherwise,\cr}\cr
B^{(\nu,1)}_{r',s'}(0,q)=&\cases{1 &if $r'=s'=0$\cr
{1\over 2} &if $r'=s'\ge 1$ or $s'=r'+1$\cr
0 &otherwise,\cr}\cr
B^{(\nu,1)}_{r',s'}(1,q)=&\cases{1+q &if $r'=s'=0$\cr
1+{1\over 2}q &if $s'=1$, $r'=0$\cr
{1\over 2}+{1\over 2}q &if $r'=s'\ge 1$ or $s'=r'+1\ge 2$\cr
1 &if $s'=0$, $r'=1$\cr
{1\over 2} &if $r'=s'+1\ge 2$ or $s'=r'+2$ \cr
0 &otherwise,\cr}\cr
B^{(\nu,-1)}_{r',s'}(0,q)=&\cases{0 &if $r'<s'$\cr
(-1)^{r'+s'} &if $r'\ge s'$,\cr}\cr
B^{(\nu,-1)}_{r',s'}(1,q)=&\cases{1 &if $r'=s'=0$ or $s'=r'+1$\cr
(-1)^{r'+s'+1} &if $r'>s'$\cr
0 &otherwise.}\cr}}
Equations~\polyidenta--\polyidentc~may be readily verified using these
expressions.
The character identities~\bfns--\bfrm~will follow from the $L\rightarrow\infty$
limit of the polynomial identities~\polyidenta--\polyidentc~thanks
to~\pcf~and~\pcb.

We close this presentation of results and methods with several remarks.
First, attention should be drawn to the presence in the fermionic forms of
solutions~\extra~to the system~\countg~with negative values for
$m_{\nu}$. This is the first time such solutions have been explicitly
encountered, but it is expected that they will also be found in
other nonunitary
models such as $M(p,p')$ for $p+1\neq p'$. Secondly, we direct attention to
the occurrence of linear combinations  in the $R^{\pm}$ sectors~\polyidentb
--\polyidentc. Such linear combinations have been seen in
several other situations and are presumably a generic feature of Fermi/Bose
correspondences although for the unitary model $M(p,p+1)$ the Bose and
Fermi polynomials appear only singly.
We also remark on the crucial role played by the fact
that there are many different polynomials which ``finitize'' the
same fermionic
character. This is a general feature which, for example, occurs in the proof
of the identities of the nonunitary $M(p,p')$ minimal model ~\rbm.

Finally, we comment that the Fermi and Bose polynomials and the corresponding
recursion relations given here are not particularly unique. As an example,
for $R^+$ with $s'=0$ we have two alternative representations
\eqn\altb{B^{(\nu,1)}_{r',0}(L,q)=B1^{(\nu)}_{r'}(L,q)=B2^{(\nu)}_{r'}(L,q)}
where
\eqn\altbdef{\eqalign{B1^{(\nu)}_{r'}(L,q)=\sum_{j=-\infty}^{\infty}(-1)^j
&\bigl(q^{\nu j^2}T_1(L,2\nu j+r'+1;q^{{1\over 2}})\cr
+&q^{\nu j(j-1)+{(L-r')\over 2}}T_0(L,2\nu j+r';q^{{1\over 2}})\bigr)\cr}}
and
\eqn\albbdef{\eqalign{B2^{(\nu)}_{r'}(L,q)=
\sum_{j=-\infty}^{\infty}(-1)^j q^{\nu j^2}&\biggl[\sum_{i=r'+1}^{\nu-1}
(-1)^{i+1+r'}T_1(L+1,2\nu j+i;q^{{1\over 2}})\cr
+&\sum_{i=r'+1}^{\nu-2}(-1)^{i+1+r'}
T_1(L,2\nu j+i;q^{{1\over 2}})\biggr].\cr}}

More generally, there are systems of polynomials which
reduce to the characters in the $L\rightarrow \infty$ limit and
satisfy slightly different systems of equations from the one given
here. However, in all these cases the new polynomials may be expressed as
linear combinations of the polynomials given above.

\newsec{Proof of Fermionic Recursion Relations}

We now turn to the proof that the fermionic sums of sec.~2 defined
by~\fpoly~satisfy the recursion relations~\nseqn.
The proof is based upon the use of telescopic expansions
of products of $q$-binomial coefficients developed in ~\rberk.
In contrast with the many identities on $q$-trinomial coefficients
we shall use in the proof of the Bosonic identities the only identities
we require for the proof of the Fermionic recursion relations are the
elementary recursion relations for $q$-binomial coefficients
\eqn\qbinidenta{{n+m\atopwithdelims[] n}_q=
{n+m-1\atopwithdelims[] n}_q+q^m{n+m-1\atopwithdelims[]n-1}_q}
and
\eqn\qbinidentb{{n+m\atopwithdelims[] n}_q={n+m-1\atopwithdelims[] n-1}_q+
q^n{n+m-1\atopwithdelims[] n}_q.}
We note that in order for these two identities to be used in our proofs
without exception we need to use the definition~\qbin.

In order to give a compact proof we introduce the following symbolic
notation for fermionic sums
\eqn\fsym{q^{\phi({\vec n},{\vec A})}{\vec P\atopwithdelims\{\}
{\vec Q}}_{\cal D}=\sum_{\cal D}q^{\phi({\vec n},{\vec A})}
\prod_{j=1}^{\nu}{n_j+m_j+P_j\atopwithdelims[] n_j+Q_j}_q}
where
\eqn\phidfn{\phi({\vec n},{\vec A})={1\over 2}(n_1^2+N_2^2)+
\sum_{i=3}^{\nu}N_i^2+{\vec A}\cdot{\vec n},}
and ${\cal D}$~specifies the domain of summation variables $m_i$ and $n_i$
which are related by~\countg. In what follows we will use three domains
${\cal D}_{r',s'},{\tilde{\cal D}}_{r',s'},{\cal D}'_{r',s'}$, where

1) ${\cal D}_{r',s'}$ was defined in sec.~2 by~\extra;

2) ${\tilde{\cal D}}_{r',s'}$ is defined by~\constraint--\eqnvb~and
\eqn\moreex{m_{\nu}=-2r',~~n_{\nu-r'}=n_{\nu-r'+1}=\cdots
=n_{\nu}=0,~~n_1,n_2,\cdots ,n_{\nu-r'-1}\geq 0;}

3) ${\cal D}'_{r',s'}$ is defined the same way as ${\cal D}_{r',s'}$ except
that $n_{\nu-r'-1}\geq -1$ (whereas it was $n_{\nu-r'-1}\geq 0$ for
${\cal D}_{r',s'}$).

In terms of this notation we write the fermionic polynomials for arbitrary
${\vec A}$
\eqn\arbf{F_{r',s'}^{\nu}(L,{\vec A},q)=q^{\phi({\vec n},{\vec A})}
{0 \atopwithdelims\{\} 0}_{{\cal D}_{r',s'}}.}

To avoid bulky formulas we find it convenient to use shorthand notations
\eqn\shrthnda{F_{r'}(L)\equiv F_{r',s'}^{\nu}(L,{\vec A},q)}
\eqn\shrthndb{\phi({\vec n})\equiv \phi({\vec n},{\vec A})}
throughout the rest of this section.

All the equations of~\nseqn~are special cases of the following
set of recursion relations for $\nu\geq 3$
\eqn\genrra{F_0(L)=F_1(L-1)+(q^{L-{1\over 2}+\alpha}+1)
F_0(L-1)+(q^{L-1+\beta}-1)F_0(L-2)}
\eqn\genrrb{\eqalign{F_{r'}(L)&=F_{r'-1}(L-1)+F_{r'+1}(L-1)\cr
&+q^{L-{1\over 2}+\alpha}F_{r'}(L-1)+(q^{L-1+\beta}-1)F_{r'}(L-2)~{\rm
for}~1\leq r' \leq \nu-3\cr}}
and
\eqn\genrrc{F_{\nu-2}(L)=F_{\nu-3}(L-1)+q^{L-{1\over 2}+\alpha}
F_{\nu-1}(L-1)+q^{L-1+\beta}F_{\nu-1}(L-2)}
where
\eqn\atilde{\eqalign{\alpha &=A_2-{\tilde a}_1^{(s')}\cr
\beta &=A_1+\alpha \cr
{\tilde a}_i^{(s')}&=\sum_{j=i}^{\nu}a^{(s')}_j=\delta_{s',\nu-1}\delta_{i,1}
+{1\over 2}\theta(\nu-s'<i)\cr}}
and $\theta(a<b)=1$ if $a<b$ and $0$ otherwise.

\noindent
When $\nu=2$ the equation~\nutns~follows from the single equation
\eqn\gennut{F_0(L)=(1+q^{L-{1\over2}+\alpha})
F_0(L-1)+q^{L-1+\beta}F_0(L-2)}
We will find that in order for~\genrra--\genrrc~to hold
${\vec A},{\vec{\tilde a}}^{(s')}$ should satisfy
\eqn\aconst{A_{i+1}-A_i=2{\tilde a}_{i+1}^{(s')},~~{\rm for}~2\leq i\leq\nu-1.}
As a consequence of~\aconst~only $A_1$ and $A_2$ may be specified
independently of the inhomogeneous vector ${\vec a}^{(s')}.$

\noindent
Making use of~\qf,~\lf,~\atilde~and
\eqn\form{m_1=N_2-n_1}
one verifies that for ${\vec A}$ defined by~\aconst~with
\eqn\nsra{A_1=-{n\over 2},~~A_2={n\over 2}+\delta_{s',\nu-1};~~~n=0,\pm 1,}
$\alpha\rightarrow {n\over 2},\beta\rightarrow 0,\phi({\vec n})\rightarrow Qf+
Lf_{n,s'}$ and therefore the fermionic forms~\arbf~and recursion relations
{}~\genrra--\genrrc~reduce to~\fpoly~and~\nseqn.

Let us denote the set of solutions of~\countg~with the inhomogeneous
vector~\indefn~as $\{{\vec n},{\vec m}\}_{L,r',s'}.$
Then, if we define vectors ${\vec e}_l$ and ${\vec E}_{l,k}$ by
\eqn\dfe{({\vec e}_l)_i=\delta_{l,i}~~~~{\vec E}_{l,k}=-\sum_{i=l}^k{\vec e}_i}
we may use~\countg~to verify the following relations
\eqn\mnrels{\eqalign{\{{\vec n},{\vec m}\}_{L-1,r'-1,s'}&=\{{\vec n},{\vec
m}\}_{L,r',s'}+\{0,{\vec E}_{2,\nu-r'}\}\cr
\{{\vec n},{\vec m}\}_{L-1,r',s'}&=\{{\vec n},{\vec
m}\}_{L,r',s'}+\{-{\vec e}_2,-{\vec e}_1\}\cr
\{{\vec n},{\vec m}\}_{L-2,r',s'}&=\{{\vec n},{\vec
m}\}_{L,r',s'}+\{-{\vec e}_1-{\vec e}_2,0\}\cr
\{{\vec n},{\vec m}\}_{L-2,r',s'}&=\{{\vec n},{\vec
m}\}_{L,r',s'}+\{{\vec e}_{\nu-r'-1}-{\vec e}_{\nu-r'},2{\vec
E}_{2,\nu-r'-1}\}~{\rm for}~\nu-r'\geq 3\cr
\{{\vec n},{\vec m}\}_{L-1,r'+1,s'}&=\{{\vec n},{\vec
m}\}_{L,r',s'}+\{{\vec e}_{\nu-r'-1}-{\vec e}_{\nu-r'},{\vec
E}_{2,\nu-r'-1}\} ~{\rm for}~\nu-r'\geq 3\cr.}}
Furthermore if we recall
\eqn\recall{\eqalign{L&=n_1+a_1+\sum_{i=2}^{\nu}(2i-3)(n_i+a_i)+m_{\nu}\cr
L&=n_1+N_2+m_2+{\tilde a}_1\cr
m_i&=2\sum_{l=i+1}^{\nu}(N_l+{\tilde a}_l)+m_{\nu};~~~i\geq 2 \cr}}
and use~\aconst, we may verify the following identities for
$\phi({\vec n},{\vec A})$
\eqn\phirel{\eqalign{&\phi({\vec n})+n_1+m_2=\phi({\vec n}-{\vec e}_2)+
L-{1\over 2}+\alpha\cr
&\phi({\vec n})+m_2=\phi({\vec n}-{\vec e}_1-{\vec e}_2)+L-1+\beta\cr
&[\phi+m_l]({\vec n}-{\vec e}_{l-1}+{\vec e}_l+{\vec e}_{\nu-r'-1}-
{\vec e}_{\nu-r'})={\tilde{\phi}}({\vec n})+m_{l-1}-1,
{}~~{\rm for}~3\leq l \leq \nu-r'\cr
&{\rm with}~~{\tilde{\phi}}({\vec n})\equiv\phi({\vec n}+{\vec e}_{\nu-r'-1}-
{\vec e}_{\nu-r'}).\cr}}
Then from~\mnrels~and~\phirel~we obtain the following expressions:
\eqn\feqna{ F_{r'}(L)=q^{\phi({\vec n})}
{0\atopwithdelims\{\}0}_{{\cal D}_{r',s'}}}

\eqn\feqnb{ F_{r'-1}(L-1)=q^{\phi({\vec n})}
{{\vec E}_{2,\nu-r'} \atopwithdelims\{\} 0}_{{\cal D}_{r',s'}}-{\cal B}}

\eqn\feqnc{q^{L-{1\over 2}+\alpha} F_{r'}(L-1)=q^{\phi({\vec n})+n_1+m_2}
{{\vec E}_{1,2}\atopwithdelims\{\} -{\vec e}_2}_{{\cal D}_{r',s'}}}

\eqn\feqnd{q^{L-1+\beta} F_{r'}(L-2)=q^{\phi({\vec n})+m_2}
{{\vec E}_{1,2} \atopwithdelims\{\} {\vec E}_{1,2}}_{{\cal D}_{r',s'}}}

\eqn\feqne{F_{r'}(L-2)=q^{{\tilde\phi}({\vec n})}{2{\vec E}_{2,\nu-r'-1}+
{\vec e}_{\nu-r'-1}-{\vec e}_{\nu-r'} \atopwithdelims\{\}{\vec e}_{\nu-r'-1}-
{\vec e}_{\nu-r'}}_{{\cal D}'_{r',s'}}}

\eqn\feqnf{F_{r'+1}(L-1)=q^{{\tilde\phi}({\vec n})}
{{\vec E}_{2,\nu-r'-1}+{\vec e}_{\nu-r'-1}-{\vec e}_{\nu-r'}
 \atopwithdelims\{\}{\vec e}_{\nu-r'-1}-
{\vec e}_{\nu-r'}}_{{\cal D}'_{r',s'}}+{\cal B}}
where
\eqn\bt{{\cal B}\equiv q^{\phi({\vec n)}}\theta(s'>r'){{\vec E}_{2,\nu-r'-1}
\atopwithdelims\{\} 0}_{\tilde {\cal D}_{r',s'}}}
and $\theta(s'>r')=1$ if $s'>r'$ and $0$ otherwise. We note that the term
$\cal B$ arises because in general ${\cal D}_{r',s'}\neq{\cal D}_{r'\pm 1,s'}.$

The method we use to prove~\genrra--\genrrc~is the telescopic expansion
technique of ~\rberk~ which is based on the following two identities
which follow from~\qbinidenta:

1)Telescopic expansion from right to left
\eqn\telrl{{{\vec P}\atopwithdelims\{\}{\vec Q}}={{\vec P}+{\vec E}_{l,k}
\atopwithdelims\{\}{\vec Q}}+\sum_{i=l}^{k}q^{m_i+P_i-Q_i}{{\vec P}+
{\vec E}_{l,i}\atopwithdelims\{\}{\vec Q}-{\vec e}_i};}

2)Telescopic expansion from left to right
\eqn\tellr{{{\vec P}\atopwithdelims\{\}{\vec Q}}={{\vec P}+{\vec E}_{l,k}
\atopwithdelims\{\} {\vec Q}}+\sum_{i=l}^k q^{m_i+P_i-Q_i}{{\vec P}+
{\vec E}_{i,k}\atopwithdelims\{\}{\vec Q}-{\vec e}_i}.}

The proof of~\genrra~will follow from~\genrrb~with the definition
$F_{-1}(L)\equiv F_0(L).$
To prove~\genrrb~we begin by applying the right to left
telescopic expansion~\telrl~to $F_r(L)$ to obtain
\eqn\peqna{F_{r'}(L)=q^{\phi({\vec n})}{{\vec E}_{2,\nu-r'}
\atopwithdelims\{\} 0}_{{\cal D}_{r',s'}}+
\sum_{l=2}^{\nu-r'}q^{\phi({\vec n})+m_l}{{\vec E}_{2,l}
\atopwithdelims\{\}-{\vec e}_l}_{{\cal D}_{r',s'}}}
and then further expand the term in the sum with $l=2$ using~\qbinidentb~to get
\eqn\peqnb{\eqalign{F_{r'}(L)&=q^{\phi({\vec n})}
{{\vec E}_{2,\nu-r'}\atopwithdelims\{\} 0}_{{\cal D}_{r',s'}}-{\cal B}\cr
&+q^{\phi({\vec n})+n_1+m_2}{{\vec E}_{1,2}\atopwithdelims\{\}-
{\vec e}_2}_{{\cal D}_{r',s'}}+q^{\phi({\vec n})+m_2}{{\vec E}_{1,2}
\atopwithdelims\{\}{\vec E}_{1,2}}_{{\cal D}_{r',s'}}+Z}}
where
\eqn\peqc{Z\equiv\sum_{l=3}^{\nu-r'}q^{\phi({\vec n})+m_l}{{\vec E}_{2,l}
\atopwithdelims\{\}-{\vec e}_l}_{{\cal D}_{r',s'}}+{\cal B}.}
Then making use of~\feqnb--\feqnd~we find
\eqn\peqd{F_{r'}(L)-F_{r'-1}(L-1)-q^{L-{1\over 2}+\alpha}
F_{r'}(L-1)-q^{L-1+\beta}F_{r'}(L-2)=Z.}
We then change the summation variables in the $l^{th}$ term in the
expansion of $Z$ as
\eqn\peqe{{\vec n}\rightarrow{\vec n}-{\vec e}_{l-1}+{\vec e}_{l}+
{\vec e}_{\nu-r'-1}-{\vec e}_{\nu-r'}{\rm~~~~~~~~~for}~3\leq l\leq\nu-r'}
where we note that this change depends on $l$ and sends the domain
${\cal D}_{r',s'}$ to ${\cal D}'_{r',s'}.$ Thus, making use of~\phirel,
we obtain
\eqn\peqf{Z=\sum_{l=2}^{\nu-r'-1}q^{{\tilde\phi}({\vec n})+m_l-1}
{{\vec E}_{2,\nu -r'-1}+{\vec E}_{l,\nu-r'-1}+{\vec e}_{\nu-r'-1}-
{\vec e}_{\nu-r'}\atopwithdelims\{\} -{\vec e}_l+{\vec e}_{\nu-r'-1}-
{\vec e}_{\nu-r'}}_{{\cal D}'_{r',s'}}+{\cal B}.}

To complete the proof we expand $F_{r'+1}(L-1)-{\cal B}$ given by~\feqnf~using
the left to right telescopic expansion~\tellr~as
\eqn\peqg{\eqalign{F_{r'+1}(L-1)&=q^{{\tilde \phi}({\vec n})}
{{\vec E}_{2,\nu-r'-1}+{\vec e}_{\nu-r'-1}-{\vec e}_{\nu-r'}
\atopwithdelims\{\}
{\vec e}_{\nu-r'-1}-{\vec e}_{\nu-r'}}_{{\cal D}'_{r',s'}}+{\cal B}\cr
&=q^{{\tilde \phi}({\vec n})}{2{\vec E}_{2,\nu-r'-1}+{\vec e}_{\nu-r'-1}-
{\vec e}_{\nu-r'}\atopwithdelims\{\} {\vec e}_{\nu-r'-1}-
{\vec e}_{\nu-r'}}_{{\cal D}'_{r',s'}}\cr
&+\sum_{l=2}^{\nu-r'-1}q^{{\tilde\phi}({\vec n})+m_l-1}{{\vec E}_{2,\nu-r'-1}+
{\vec E}_{l,\nu-r'-1}+{\vec e}_{\nu-r'-1}-{\vec e}_{\nu-r'}\atopwithdelims\{\}
-{\vec e}_{l}+{\vec e}_{\nu-r'-1}-{\vec e}_{\nu-r'}}_{{\cal D}'_{r',s'}}
+{\cal B}.}}

Thus comparing the right hand side of~\peqg~with~\feqne~and~\peqf~we obtain
\eqn\peqh{F_{r'+1}(L-1)-F_{r'}(L-2)=Z}
and hence, the desired result~\genrrb~follows from comparing~\peqd~and~\peqh.

It remains to prove~\genrrc. To do this we expand $F_{\nu-2}(L)$ as
\eqn\peqi{F_{\nu-2}(L)=q^{\phi({\vec n})}{{\vec E}_{2,2}
\atopwithdelims\{\} 0}_{{\cal D}_{\nu-2,s'}}+q^{\phi({\vec n})+n_1+m_2}
{{\vec E}_{1,2}\atopwithdelims\{\}-{\vec e}_2}_{{\cal D}_{\nu-2,s'}}+
q^{\phi({\vec n})+m_2}{{\vec E}_{1,2}\atopwithdelims\{\}
{\vec E}_{1,2}}_{{\cal D}_{\nu-2,s'}}}
from which~\genrrc~follows upon using~\feqnb--\feqnd.

The proof of the equation~\gennut~for $\nu=2$ is completely analogous
to the proof of~\genrrc~and will be omitted.

We close this section with a few remarks. The major new feature of
this derivation which did not occur in ~\rberk~ is the occurrence of the
extra terms~\extra~in the allowed range of solutions  ${\cal D}_{r',s'}$
of the constraint equations~\countg. These terms are forced upon us by
the necessity of using the recursion relation~\qbinidenta~for the
case $m=n=0$ and is what requires us to keep track of the three
different domains of definitions ${\cal D}_{r',s'},$ ${\cal D}'_{r',s'}$ and
${\tilde{\cal D}}_{r',s'}$ and the resulting boundary terms ${\cal B}.$ This
complicates the presentation, but since none of these terms make an
explicit contribution to the equations we advise the reader to ignore
them on first reading. Clearly, the method used can be extended to the
general case where  ${\vec{\tilde a}}^{(s')}$ and ${\vec A}$ are subject only
to~\aconst. We plan to discuss this in a separate publication.

\newsec{Proof of Bosonic Recursion Relations}

Our proof that the bosonic forms~\bosepolyns--\bosepolyr~satisfy the
recursion relations~\nseqn~relies on various identities satisfied by the
$q$-trinomial coefficients. Some of these have appeared previously in the
literature~\rab--\randc~and some occur in this proof for the first time.
For clarity we will first list all the identities we shall require and
relegate the proofs of the new ones to the Appendix. We will then use these
identities to verify the bosonic form of the recursion relations. The three
distinct cases will be considered in separate subsections for $\nu\geq 3.$
The special case $\nu=2$ is easily treated by the same methods but the
proof will be omitted.

\subsec{Identities of $q$-trinomials}

In the course of our proofs we will need several identities
satisfied by the $q$-trinomials. These identities are of three types:

A) Pascal triangle identities which are nontrivial for $q=1:$
\eqn\identa{\eqalign{T_{-n}(L,A;q^{{1\over 2}})=&
T_{-n}(L-1,A+1;q^{{1\over 2}})+T_{-n}(L-1,A-1;q^{{1\over 2}})\cr
+&q^{L-{1-n\over 2}}T_{-n}(L-1,A;q^{{1\over 2}})+(q^{L-1}-1)
T_{-n}(L-2,A;q^{{1\over 2}}),\cr}}
\eqn\identf{\eqalign{T_1&(L,A;q^{{1\over 2}})-T_1(L-1,A;q^{{1\over 2}})\cr
&=q^{L+A\over
2}T_0(L-1,A+1;q^{{1\over 2}})+q^{L-A\over2}T_0(L-1,A-1;q^{{1\over 2}}),\cr}}

B) Identities derivable from the Pascal triangle identities (for $q=1$):
\eqn\identb{\eqalign{T_0&(L,A;q^{1\over 2})-T_0(L-1,A-1;q^{1\over 2})\cr
&=q^{A+{1\over 2}}\bigl[T_0(L,A+1;q^{1\over 2})-
T_0(L-1,A+2;q^{1\over 2})\bigr],\cr}}
\eqn\idente{\eqalign{T_1&(L,A;q^{{1\over 2}})-T_1(L,A+1;q^{{1\over 2}})\cr
&=q^{L-A\over 2}T_0(L-1,A-1;q^{{1\over 2}})-
q^{L+A+1\over 2}T_0(L-1,A+2;q^{{1\over 2}}),\cr}}
\eqn\sam{\eqalign{T_1&(L+1,A;q^{{1\over 2}})+T_1(L,A;q^{{1\over 2}})\cr
&=T_{-1}(L,A+1;q^{{1\over 2}})+T_{-1}(L,A-1;q^{{1\over 2}})
+2T_{-1}(L,A;q^{{1\over 2}})\cr}}

C) Identities which become tautologies when $q=1:$
\eqn\identd{\eqalign{T_1&(L,A;q^{{1\over 2}})-T_1(L,A+1;q^{{1\over 2}})\cr
&=q^{L-A\over
2}T_0(L,A;q^{{1\over 2}})-q^{L+A+1\over 2}T_0(L,A+1;q^{{1\over 2}}),}}
\eqn\identi{\eqalign{T_{-1}(L,A;q^{{1\over 2}})&
-T_{-1}(L-1,A\pm 1;q^{{1\over 2}})\cr
&=q^{L\mp A\over 2}T_{0}(L,A;q^{{1\over 2}})-
q^LT_{-1}(L-1,A\pm 1;q^{{1\over 2}})\cr}}
\eqn\identj{\eqalign{q^{L\pm A\over 2}T_0(L,A;q^{{1\over 2}})&-
T_{1}(L,A;q^{{1\over 2}})\cr
&=(q^L-1)[T_{-1}(L-1,A;q^{{1\over 2}})+T_{-1}(L-1,A\mp 1;q^{{1\over 2}})]\cr}}
\eqn\identh{\eqalign{T_{0}&(L,A;q^{{1\over 2}})-T_{0}(L,A+2;q^{{1\over 2}})\cr
&=q^{L-A\over 2}T_1(L,A;q^{{1\over 2}})-q^{L+2+A\over 2}
T_1(L,A+2;q^{{1\over 2}})\cr}}

The identities~\identa~ with $n=0$ and~\identb~are needed for the proof in the
$NS$ sector. Identity~\identa~is proven in the appendix and~\identb~follows by
combining eqns. (2.26) and (2.29) of~\rab.

The identities~\identa~with $n=-1$,~\identf,\idente~and~\identd~are needed for
the $R^-$ Ramond sector.  Identity~\identf~is eqn.(2.16) of~\rab,
identity~\identd~is eqn. (2.20) of~\rab, and identity~\idente~follows from
combining~\identb~and~\identd.

The identities~\identa~with $n=1$,~\identi--\identh~are needed for the $R^+$
Ramond sector. Identity~\identi~is (2.23) of~\rab~with $B=A+1,$
identity~\identj~is obtained by combining (2.23) and (2.24) of~\rab~both
with $B=A+1$ and identity~\identh~will be proven in the appendix.
Finally, identity~\sam~is needed to establish an $R^+~-~R^-$ connection and to
derive~\tfnlim. Identity~\sam~will also be proven in the appendix.

\subsec{Proof of the generic equations for all sectors}

We separate the recursion relations~\nseqn~into two classes: the equations
for $h_0,\cdots,h_{\nu-3}$ which we call generic and the last equation
of~\nseqn~(or equivalently~\hnmo) which we call the closing equation.
The proof of the generic equations is identical for the three separate
cases of $NS$ and $R^{\pm}.$ In all cases the generic equation
follows immediately from the identity~\identa~and the fact that the
bosonic polynomials~\bosepolyns,~\bosepolyrp,~\bosepolyr~are linear
combinations of $T_{-n}$ with $n$ given by~\ndefn.
The identity~\identa~guarantees that the generic recursion relation holds for
each term separately in the sum over $j$. Consequently, these generic
equations do not determine the factors
$(-1)^j q^{\nu j^2+(s'+{1-|n| \over 2})j}$
which appear in~\bosepolyns--\bosepolyr.
These factors are determined by the closing equation and for this the three
cases need to be considered separately.

To keep notations manageable we will write $T_n(L,A)$ instead of
$T_n(L,A;q^{1\over 2})$ throughout the rest of this paper.

\subsec{Proof of the closing equation for the Neveu-Schwarz sector}

To verify the closing equation~\hnmo~for the $NS$ bosonic
polynomials~\bosepolyns~we consider
\eqn\ieqn{I_{NS}(L)=B^{(\nu,0)}_{\nu-1,s'}(L,q)-B^{(\nu,0)}_{\nu-2,s'}(L-1,q)}
and substitute~\bosepolyns~to find
\eqn\iaeqn{\eqalign{I_{NS}(L)=&\sum_{j=-\infty}^{\infty}(-1)^j
q^{\nu j^2+(s'+{1\over 2})j}\biggl( T_0(L,2\nu j +s'-\nu+1)+
T_0(L,2\nu j+s'+\nu)\cr
&-T_0(L-1,2\nu j +s'-\nu +2)-T_0(L-1,2\nu j +s' +\nu -1)\biggr).\cr}}
This does not vanish term by term under the summation sign. However, if
we first send $j\rightarrow -j$ in the first and third terms inside
of $\bigl(\cdots\bigr)$ and use~\prop~we have
\eqn\ibeqn{\eqalign{I_{NS}(L)=&\sum_{j=-\infty}^{\infty}(-1)^j
q^{\nu j^2} \biggl(q^{(s'+{1\over2})j}\bigl[T_0(L,2\nu j +\nu +s')-
T_0(L-1,2\nu j +\nu +s'-1)\bigr]\cr
&~~~~~~+q^{-(s'+{1\over 2})j}\bigl[T_0(L,2\nu j+\nu-s'-1)-
T_0(L-1,2\nu j+\nu -s'-2)\bigr]\biggr).\cr}}
In this sum the terms with $j$ and $-j-1$ cancel by use of~\identb.
Thus we have completed the verification that the $NS$ bosonic
polynomials~\bosepolyns~satisfy the recursion relations~\nseqn~with $n=0$.

\subsec{Proof of the closing equation for the $R^-$ Ramond sector}

To verify the closing equation~\hnmo~for the $R^-$ polynomials~\bosepolyr~we
consider
\eqn\grameqc{I_{R^-}(L)=B^{(\nu,-1)}_{\nu-1,s'}(L,q)
-B^{(\nu,-1)}_{\nu-2,s'}(L-1,q)}
and substitute~\bosepolyr~to find
\eqn\grameqd{\eqalign{I_{R^-}(L)=&\sum_{j=-\infty}^{\infty}(-1)^j
q^{\nu j^2+s'j}\cr
&\times \biggl(\sum_{i=-(\nu-1)}^{\nu-1}(-1)^{\nu-1+i}T_1(L,2\nu j+ s'+i)\cr
&-\sum_{i=-(\nu-2)}^{\nu-2}(-1)^{\nu-2+i}T_1(L-1,2\nu j+s'+i)\biggr).\cr}}
We now transform the summand of~\grameqd~for each $j$ by adding and
subtracting $T_1(L-1,2\nu j +s'-1+\nu)$ and regrouping terms to obtain
\eqn\grameqe{\eqalign{&\sum_{i=0}^{\nu-2}\bigl[T_1(L,2\nu j
+s'+1-\nu+2i)-T_1(L,2\nu j +s'+2-\nu+2i)\bigr]\cr
&+\bigl[T_1(L,2\nu +s'-1+\nu)-T_1(L-1,2\nu j+s'-1+\nu)\bigr]\cr
&-\sum_{i=0}^{\nu-2}\bigl[T_1(L-1,2\nu j+s'+2-\nu+2i)-T_1(L-1,2\nu
j+s'+3-\nu+2i)\bigr].\cr}}
Then we use~\idente~on the first line,~\identf~on the second line
and~\identd~on the third line and note that all terms
cancel in pairs except
\eqn\grameqf{q^{L-1+\nu\over 2}\biggl(q^{2\nu j+s'\over 2}
T_0(L-1,2\nu j+s'+\nu)+q^{-{2\nu j+s'\over 2}}T_0(L-1,2\nu j+s'-\nu)\biggr).}
Thus we have
\eqn\grameqg{\eqalign{I_{R^-}(L)=&q^{L-1+\nu\over 2}
\sum_{j=-\infty}^{\infty}(-1)^j q^{\nu j^2+s'j}\cr
&\times \bigl[q^{2\nu j + s'\over 2}T_0(L-1,2\nu j +s'+\nu)+q^{-{2\nu
j+s'\over 2}}T_0(L-1,2\nu j+s'-\nu)\bigr].\cr}}
which is seen to vanish if we replace $j$ by $j+1$ in the second of two terms
in $\bigl[\cdots\bigr]$.
Thus we have completed the verification that the $R^-$ bosonic
polynomials~\bosepolyr~satisfy the recursion relations~\nseqn~with $n=-1.$

\subsec{Proof of the closing equation for the $R^+$ Ramond sector, $R^+-R^-$
relations}

To verify the closing equation~\hnmo~for the $R^+$ polynomials ~\bosepolyrp~we
consider
\eqn\irpd{I_{R^+}(L)=B^{(\nu,1)}_{\nu-1,s'}(L,q)-B^{(\nu,1)}_{\nu-2,s'}(L-1,q)}
and substitute~\bosepolyrp~to find
\eqn\rpone{\eqalign{I_{R^+}(L)=&\sum_{j=-\infty}^{\infty}(-1)^j
q^{\nu j^2+s' j}\biggl([T_{-1}(L,2\nu j+s'-\nu +1)-
T_{-1}(L-1,2\nu j+s'-\nu+2)]\cr
{}~~~~~+&[T_{-1}(L,2\nu j+s'+\nu)-T_{-1}(L-1,2\nu j+s'+\nu-1)]\cr
{}~~~~~+&[T_{-1}(L,2\nu j+s'-\nu)
-T_{-1}(L-1,2\nu j+s'-\nu+1)]\cr
{}~~~~~+&[T_{-1}(L,2\nu j+s'+\nu-1)-T_{-1}(L-1,2\nu j+s'+\nu-2)]\biggr).}}
We now use~\identi~on each of the four terms inside of $\bigl(\cdots\bigr)$
to obtain after regrouping
\eqn\rptwo{\eqalign{I_{R^+}(L)&=\sum_{j=-\infty}^{\infty}(-1)^j
q^{\nu j^2+s' j}\times \cr
&\biggl([q^{L-(s'-\nu+1+2 \nu j)\over 2}T_0(L,2\nu j+s'-\nu+1)+
q^{L+(s'+\nu-1+2 \nu j)\over 2}T_0(L,2\nu j+s'+\nu-1)]\cr
&-q^L[T_{-1}(L-1,2\nu j+s'-\nu+2)+T_{-1}(L-1,2\nu j+s'-\nu+1)]\cr
&-q^L[T_{-1}(L-1,2\nu j+s'+\nu-1)+T_{-1}(L-1,2\nu j+s'+\nu-2)]\cr
&+[q^{L+s'+\nu+2 \nu j\over 2}T_0(L,2\nu j+s'+\nu)+
q^{L-(s'-\nu+2\nu j)\over 2} T_0(L,2\nu j+s'-\nu)]\biggr).\cr}}
The expression in the last set of the square brackets is seen to vanish
if we take $j\rightarrow j+1$ in the second of the two terms in $[\cdots].$
Then, if we multiply both sides of~\rptwo~by $(q^L-1)$ and use~\identj~on
the contents of the second and third sets of square brackets, we obtain
\eqn\rpthre{\eqalign{(q^L-1)&I_{R^+}(L)=\sum_{j=-{\infty}}^{\infty}
q^{\nu j^2+s' j}\times \cr
&\biggl(-q^{L+s'+\nu-1+2 \nu j\over 2}T_0(L,2\nu j+s'+\nu-1)
-q^{L-(s' -\nu +1+2 \nu j)\over 2}T_0(L,2 \nu j+s'-\nu+1)\cr
{}~~~~~~&+q^LT_1(L,2\nu j+s'+\nu-1)+q^LT_1(L,2\nu j+s'-\nu+1)\biggr).\cr}}
We now let $j\rightarrow j-1$ in the first and third terms in
$\bigl(\cdots\bigr)$ to obtain the expression
\eqn\rpfour{\eqalign{(q^L-1)I_{R^+}(L)&=\sum_{j-\infty}^{\infty}
(-1)^jq^{\nu j (j-1)+s' j+{L+\nu -s'-1\over 2}}\cr
&\times \biggl(T_0(L,2\nu j+s'-\nu -1)-T_0(L,s'-\nu+1)\cr
&-q^{L-(s'-\nu-1+2\nu j)\over 2}T_1(L,2\nu j+s'-\nu-1)\cr
&+q^{L+(s'-\nu+1+2\nu j)\over 2}T_1(L,2\nu j+s'-\nu +1)\biggr)\cr}}
which vanishes term by term under the summation sign due to~\identh.
Thus we have completed the verification that the $R^+$ bosonic
polynomials~\bosepolyrp~satisfy the recursion relations~\nseqn~with $n=1.$

We conclude this section by noting the following intriguing identities
\eqn\koza{\eqalign{&F^{(\nu,-1)}_{r',s'}(L+1,q)+F^{(\nu,-1)}_{r',s'}(L,q)=
2\bigl[B^{(\nu,1)}_{r',s'}(L,q)+B^{(\nu,1)}_{r',s'+1}(L,q)\bigr];~~~
s'\neq \nu-1 \cr
&F^{(\nu,-1)}_{r',\nu-1}(L+1,q)+F^{(\nu,-1)}_{r',\nu-1}(L,q)=
2B^{(\nu,1)}_{r',\nu-1}(L,q),\cr}}
which can be easily proven with the help of~\polyidentc~and~\sam.

\noindent
These identities reveal the intimate connection between $R^+$ and $R^-$
representations of the Ramond sector characters.

\newsec{The Indices}

In this section we turn to the objects ${\tilde F}_{s'}^{(\nu,n)}(q)$ and
${\tilde B}_{s'}^{(\nu,n)}(q)$  and prove the properties discussed in the
introduction. To this end we introduce the polynomials
${\tilde F}^{(\nu,n)}_{r',s'}(L,q)$ as
\eqn\indf{{\tilde F}_{r',s'}^{(\nu,n)}(L,q)=\sum_{{\cal D}_{r',s'}}
(-1)^{m_1}q^{Qf+Lf_{n,s'}}\prod_{j=1}^{\nu}{n_j+m_j\atopwithdelims[] n_j}_q;
{}~~~~~n=0,\pm 1,~~~r'=0,1,\cdots,\nu-2}
where $Qf,~Lf_{n,s'},$ and ${\cal D}_{r',s'}$ are defined
in~\qf,~\lf~and~\extra.
One can easily establish
\eqn\bambi{\lim_{L\rightarrow\infty}{\tilde F}_{r',s'}^{(\nu,n)}(L,q)=
{\tilde F}^{(\nu,n)}_{s'}(q)}
and
\eqn\thumper{{\tilde F}_{r',s'}^{(\nu,0)}(L,q)=
F^{(\nu,0)}_{r',s'}(L,qe^{2\pi i})}
which hold for all $r'.$

It is straightforward to repeat the analysis carried out in secs.~3 and 4 to
prove recursion relations for ${\tilde F}_{r',s'}^{(\nu,n)}(L,q)$
\eqn\tgenrra{{\tilde F}^{(\nu,n)}_{0,s'}(L,q)={\tilde F}^{(\nu,n)}_{1,s'}
(L-1,q)+(1-q^{L-{1-n\over 2}})
{\tilde F}^{(\nu,n)}_{0,s'}(L-1,q)+(q^{L-1}-1)
{\tilde F}^{(\nu,n)}_{0,s'}(L-2,q),}
\eqn\tgenrrb{\eqalign{{\tilde F}^{(\nu,n)}_{r',s'}(L,q)&=
{\tilde F}^{(\nu,n)}_{r'-1,s'}(L-1,q)+
{\tilde F}^{(\nu,n)}_{r'+1,s'}(L-1,q)\cr
&-q^{L-{1-n\over 2}}{\tilde F}^{(\nu,n)}_{r',s'}(L-1,q)+
(q^{L-1}-1){\tilde F}^{(\nu,n)}_{r',s'}(L-2,q)
{}~{\rm for}~1\leq r'\leq\nu-3,\cr}}
\eqn\tgenrrc{{\tilde F}^{(\nu,n)}_{\nu-2,s'}(L,q)=
{\tilde F}^{(\nu,n)}_{\nu-3,s'}(L-1,q)-q^{L-{1-n\over 2}}
{\tilde F}^{(\nu,n)}_{\nu-2,s'}(L-1,q)+q^{L-1}
{\tilde F}^{(\nu,n)}_{\nu-2,s'}(L-2,q);}
with $\nu\geq 3$,
\eqn\tgennut{{\tilde F}^{(2,n)}_{0,s'}(L,q)=(1-q^{L-{1-n\over2}})
{\tilde F}^{(2,n)}_{0,s'}(L-1,q)+q^{L-1}
{\tilde F}^{(2,n)}_{0,s'}(L-2,q)}
and identities
\eqn\ariel{{\tilde F}_{r',s'}^{(\nu,0)}(L,q)=
{\tilde B}^{(\nu,0)}_{r',s'}(L,q)}
\eqn\sebastian{{\tilde B}_{r',s'}^{(\nu,1)}(L,q)=
{\tilde F}^{(\nu,1)}_{r',s'}(L,q)
-{\tilde F}^{(\nu,1)}_{r',s'-1}(L,q),~~~~~s'\neq 0}
where
\eqn\tildepolyns{\eqalign{{\tilde B}^{(\nu,0)}_{r',s'}(L,q)=
(-1)^{L+s'+r'}&\sum_{j=-\infty}^{\infty} q^{\nu
j^2+(s'+{1\over 2})j}\biggl(T_0(L,2\nu j+s'-r')\cr
&-T_0(L,2\nu j +s'+1+r')\biggr)}}
and
\eqn\btdefnpoly{\eqalign{{\tilde B}_{r',s'}^{(\nu,1)}(L,q)=(-1)^{L+r'+s'+1}&
\sum_{j=-\infty}^{\infty}q^{\nu j^2+s'j}\cr
&\times \biggl(T_{-1}(L,2\nu j+s'+r'+1)-T_{-1}(L,2\nu j+s'-r')\cr
&+T_{-1}(L,2\nu j+s'+r')-T_{-1}(L,2\nu j+s'-r'-1)\biggr)\cr.}}
Identity~\tildepolyns~could have been proven directly by simply replacing
$q^{{1\over 2}}$ with $-q^{{1\over 2}}$ in~\polyidenta~and then using
{}~\tnsym~and~\thumper. To avoid confusion we want to stress that
${\tilde B}^{(\nu,1)}_0(q)$ defined by~\tbwit~is not $L\rightarrow\infty$
limit of ${\tilde B}^{(\nu,1)}_{r',0}(L,q)$.

If we let $L\rightarrow\infty$ in~\sebastian and apply the limiting
formulas~\tzlim~
with $n=1$ and~\bambi~we obtain
\eqn\deva{{\tilde F}_{s'}^{(\nu,1)}(q)={\tilde F}_{s'-1}^{(\nu,1)}(q),~~~~~
s'\neq 0,}
i.e. ${\tilde F}_{s'}^{(\nu,1)}(q)$ does not depend on $s'$. In fact,
\eqn\hombre{{\tilde F}_{s'}^{(\nu,1)}(q)=1}
as stated in the introduction~\exramsec. To see this, we rearrange~\tgenrra-
\tgennut~in the following fashion (suppressing the argument $q$ for
compactness):

1) For $\nu=2$
\eqn\indpra{{\tilde F}^{(2,1)}_{0,s'}(L)+q^{L}
{\tilde F}^{(2,1)}_{0,s'}(L-1)=
{\tilde F}^{(2,1)}_{0,s'}(L-1)+q^{L-1}{\tilde F}^{(2,1)}_{0,s'}(L-2)}

2) For $\nu\geq 3$
\eqn\indprc{\eqalign{&{\tilde F}^{(\nu,1)}_{0,s'}(L)+
q^L{\tilde F}^{(\nu,1)}_{0,s'}(L-1)=
{\tilde F}^{(\nu,1)}_{0,s'}(L-1)+{\tilde F}^{(\nu,1)}_{1,s'}(L-1)+
(q^{L-1}-1){\tilde F}^{(\nu,1)}_{0,s'}(L-2),\cr
&{\tilde F}^{(\nu,1)}_{r',s'}(L)+q^L{\tilde F}^{(\nu,1)}_{r',s'}(L-1)-
{\tilde F}^{(\nu,1)}_{r'-1,s'}(L-1)=
{\tilde F}^{(\nu,1)}_{r'+1,s'}(L-1)+(q^{L-1}-1)
{\tilde F}^{(\nu,1)}_{r',s'}(L-2)\cr
&~~~~~~~~~~~~~~~~~~~~~~~~~~~~~~~~~~~~~~~~~~~~~~{\rm for}~1\leq r'\leq\nu-3,\cr
&{\tilde F}^{(\nu,1)}_{\nu-2,s'}(L)+q^L
{\tilde F}^{(\nu,1)}_{\nu-2,s'}(L-1)-
{\tilde F}^{(\nu,1)}_{\nu-3,s'}(L-1)=q^{L-1}
{\tilde F}^{(\nu,1)}_{\nu-2,s'}(L-2).\cr}}
We add together the $\nu-1$ equations to find
\eqn\indprd{\eqalign{\sum_{r'=0}^{\nu-2}&\bigl[
{\tilde F}^{(\nu,1)}_{r',s'}(L,q)+q^{L}{\tilde F}^{(\nu,1)}_{r',s'}(L-1,q)
\bigr]-\sum_{r'=0}^{\nu-3}{\tilde F}^{(\nu,1)}_{r',s'}(L-1,q)\cr
&=\sum_{r'=0}^{\nu-2}\bigl[{\tilde F}^{(\nu,1)}_{r',s'}(L-1,q)+
q^{L-1}{\tilde F}^{(\nu,1)}_{r',s'}(L-2,q)
\bigr]-\sum_{r'=0}^{\nu-3}{\tilde F}^{(\nu,1)}_{r',s'}(L-2,q).\cr}}
The above is of the form $I(L)=I(L-1).$ Thus both sides are separately equal
to a constant independent of $L$ which by evaluation for small $L$ is
found to be $1$ and hence
\eqn\indpre{\sum_{r'=0}^{\nu-2}\bigl[{\tilde F}^{(\nu,1)}_{r',s'}(L,q)+q^{L}
{\tilde F}^{(\nu,1)}_{r',s'}(L-1,q)\bigr]-\sum_{r'=0}^{\nu-3}
{\tilde F}^{(\nu,1)}_{r',s'}(L-1,q)=1.}
Taking~\bambi~into account we may send $L\rightarrow \infty$ in~\indpre~to
derive
\eqn\indprf{{\tilde F}_{s'}^{(\nu,1)}(q)=\lim_{L\rightarrow\infty}
{\tilde F}^{(\nu,1)}_{r',s'}(L,q)=1}
which proves~\exramsec~of the introduction.

When $\nu=2$  there is yet another bosonic companion of
${\tilde F}_{0,s'}^{(2,1)}(L,q),~~s'=0,1$. Indeed, in this case~\indpre~becomes
a simple first order difference equation
\eqn\indpret{{\tilde F}^{(2,1)}_{0,s'}(L,q)+
q^L{\tilde F}^{(2,1)}_{0,s'}(L-1,q)=1.}
By direct evaluation, one finds boundary conditions for~\indpret
\eqn\boundt{{\tilde F}^{(2,1)}_{0,s'}(s',q)=1,~~~~~s'=0,1.}
It is now trivial to solve~\indpret~and~\boundt~to obtain
\eqn\trpnut{{\tilde F}^{(2,1)}_{0,s'}(L,q)=\sum_{j=0}^{L-s'}(-1)^j
q^{Lj-{j(j-1)\over 2}}.}
{}From this the limit~\indprf~is immediate.

\noindent
Next let us consider ${\tilde F}_{0,s'}^{(2,-1)}(L,q),~s'=0,1$. If we define
\eqn\nutrmtwo{X_{s'}(L,q)={\tilde F}_{0,s'}^{(2,-1)}(L,q)-
{\tilde F}_{0,s'}^{(2,-1)}(L-1,q),}
then the second order difference equation~\tgennut~can be rewritten in the
first order form
\eqn\nutrmthree{X_{s'}(L,q)=-q^{L-1}X_{s'}(L-1,q).}
This is easily solved to get
\eqn\nutrmfour{X_{s'}(L,q)=(-1)^{L-1}q^{L(L-1)\over 2}X_{s'}(1,q)}
where $X_0(1,q)=-X_1(1,q)=-1$.
Then, since
\eqn\nutrfive{{\tilde F}_{0,0}^{(2,-1)}(0,q)={\tilde F}_{0,1}^{(2,-1)}(1,q)=1}
we obtain from~\nutrmtwo~and~\nutrmfour
\eqn\nutrmsix{\eqalign{{\tilde F}_{0,0}^{(2,-1)}(L,q)&=
1+\sum_{j=1}^L(-1)^j q^{{j(j-1)\over 2}}\cr
{\tilde F}_{0,1}^{(2,-1)}(L,q)&=1-\sum_{j=2}^L(-1)^j q^{{j(j-1)\over 2}}.\cr}}
{}From this we note that
\eqn\nutrmseven{{\tilde F}_{0,1}^{(2,-1)}(L,q)+{\tilde F}_{0,0}^{(2,-1)}(L,q)
=1.}
The equality with the false theta functions~\tfnlim~is easily
established by letting $L\rightarrow\infty$ in~\nutrmsix~and~\nutrmseven.

In fact, equation~\nutrmseven~can be generalized to
\eqn\xray{1=\sum_{s'=0}^{\nu-1}{\tilde F}_{r',s'}^{(\nu,-1)}(L,q).}
To prove~\xray~it is sufficient to notice that a constant is always a
solution to~\tgenrra--\tgenrrc~with $n=-1$ and that~\xray~holds true for
$L=0,1$. Letting $L\rightarrow\infty$ in~\xray~one recovers~\tbfrmt,~\tblim~for
$s'=0$.

To verify~\tbfrmt--\tblim~in general, we will need the following
analogue of~\koza
\eqn\kot{{\tilde F}^{(\nu,-1)}_{r',s'}(L+1,q)-{\tilde F}^{(\nu,-1)}_{r',s'}
(L,q)={\tilde B}^{(\nu,1)}_{r',s'+1}(L,q)-{\tilde B}^{(\nu,1)}_{r',s'}(L,q).}
which is proven by observing that lhs and rhs  satisfy the same
equations~\tgenrra,~\tgenrrb~with $n=1$ and that~\kot~holds true for $L=0,1$.
Then we use~\kot~along with
\eqn\iza{{\tilde F}^{(\nu,-1)}_{r',s'}(L=0,q)=\delta_{r',s'}}
to find the bosonic companion of ${\tilde F}^{(\nu,-1)}_{r',s'}(L,q)$
\eqn\koshka{{\tilde F}^{(\nu,-1)}_{r',s'}(L+1,q)=
\sum^L_{l=0}\bigl\{{\tilde B}^{(\nu,1)}_{r',s'+1}(l,q)-
{\tilde B}^{(\nu,1)}_{r',s'}(l,q)\bigr\}+\delta_{r',s'}.}
To proceed further, we set $r'=0$ and send $L$ to infinity in~\koshka~and
use~\btdefnpoly~to find
\eqn\zhopa{\eqalign{{\tilde F}^{(\nu,-1)}_{s'}(q)=\lim_{L\rightarrow\infty}
{\tilde F}^{(\nu,-1)}_{0,s'}(L,q)=(-1)^{s'}
&\biggl\{\sum_{j=-\infty}^{\infty}q^{\nu j^2+s'j} g(2\nu j+s',q)\cr
&+\sum_{j=-\infty}^{\infty}q^{\nu j^2+(s'+1)j} g(2\nu j+s'+1,q)\biggr\}+
\delta_{0,s'}\cr}}
where
\eqn\defofg{g(j,q)=\sum_{l=0}^{\infty}(-1)^l\bigl[T_{-1}(l,j+1)-
T_{-1}(l,j-1)\bigr].}

The function $g(j,q)$ has the two important properties:
\eqn\odin{g(-j,q)=-g(j,q);~~~~j\in Z}
and
\eqn\dva{g(j,q)+g(j+1,q)=-\delta_{j,0};~~~~j\in Z,~j\geq 0.}
Formula~\odin~is a simple consequence of~\prop~and formula~\dva~is
proven in the appendix. Clearly, equations~\odin~and~\dva~specify $g(j)$
uniquely as
\eqn\troyak{g(j)=\cases{(-1)^j {\rm sign}(j);& $j\neq 0$ \cr
0;& $j=0$. \cr}}
Combining~\troyak~and~\zhopa~we obtain
\eqn\pizda{{\tilde F}_{s'}^{(\nu,-1)}(q)=I^{(\nu)}_{s'}(q)
-I^{(\nu)}_{s'+1}(q)}
with $I^{(\nu)}_{s'}(q)$ defined by~\false. Thus, we completed the proof of
{}~\tfnlim.

To the best of our knowledge, $q$-trinomial representation~\koshka~of the
"truncated" false theta function has never appeared in the literature before.

We conclude this section with a derivation of the $q\rightarrow 1^-$
limit~\limfal~of the false theta function~\false~$I_{s'}^{(\nu)}(q)$
given in the introduction. To this end we rewrite the sum in~\false~as
\eqn\falone{I_{s'}^{(\nu)}(q)=1+\sum_{j=1}^{\infty}e^{-\nu x^2(j)}
\bigl(e^{-s'x(j)(|\ln q |)^{{1\over 2}}}-e^{s'x(j)(|\ln q |)^{{1\over 2}}}
\bigr)}
where
\eqn\faltwo{x(j)=j (|\ln q|)^{{1\over 2}}.}

\noindent
As $q\rightarrow 1^-$ the rhs of~\falone~is dominated by large $j$ terms and,
as a result, can be approximated by an integral
\eqn\falthree{I_{s'}^{(\nu)}(q)\sim 1+\int_0^{\infty}
{dx\over(|\ln q|)^{{1\over 2}}}e^{-\nu x^2}(e^{-s'x
(|\ln q|)^{{1\over 2}}}-e^{s'x(|\ln q |)^{{1\over 2}}}).}
Expanding
\eqn\falfour{e^{-s'x(|\ln q|)^{{1\over 2}}}-e^{s'x(|\ln q|)^{{1\over 2}}}=
-2s'x(\ln q |)^{{1\over 2}}+O(\ln q)}
we find the limit
\eqn\falfive{\eqalign{\lim_{q\rightarrow 1}I_{s'}^{(\nu)}(q)&
=1-2s'\int_{0}^{\infty}x e^{-\nu x^2} dx\cr
&=1-{s'\over\nu}}.}
The formula above is the result~\limfal~we set out to obtain.

\newsec{Duality $q\rightarrow q^{-1}$}

The bosonic and fermionic polynomials given in sec.~2 reduce to the characters
of the $SM(2,4\nu)$ superconformal model as $L\rightarrow \infty$ when $q <1.$
However, when $q > 1$ it is also possible to take the $L\rightarrow\infty$
limit after removing a suitable power $q.$ We show here that this leads to the
linear combinations of the characters of the minimal model $M(2\nu-1, 4\nu)$
where we recall that for all models $M(p,p')$ the bosonic form of the
characters (normalized to one at $q=0$) is~\rrc
\eqn\roca{\chi_{r,s}^{(p,p')}(q)=\chi_{p-r,p'-s}^{(p,p')}(q)=
{1\over (q)_{\infty}}\sum_{j=-\infty}^{\infty}(q^{j(jpp'+rp'-sp)}-
q^{(jp+r)(jp'+s)}).}

We study the region $q>1$ by making the dual transformation
$q\rightarrow q^{-1}$ in the bosonic/fermionic polynomials. It is worth
mentioning that this operation has a direct physical meaning: it transforms
particles into holes and vice-versa.

We use the definition~\tndf~to express the dual polynomials in terms of
${L,A-n; q\choose A}_2$:

1) in $NS$ as
\eqn\dualns{\eqalign{q^{{L^2\over 2}}&B_{r',s'}^{(\nu,0)}(L,q^{-1})
=\sum_{j=-\infty}^{\infty}(-1)^j q^{-\nu j^2-(s'+{1\over 2})j}\cr
&\biggl(q^{(2\nu j +s'-r')^2\over 2}{L,2\nu j+s'-r';q\choose 2\nu j+s'-r'}_2+
q^{(2\nu j+s'+1+r')^2\over 2}
{L,2\nu j+s'+r'+1;q\choose 2\nu j+r'+1}_2\biggr)\cr}}

2) in $R^-$ as
\eqn\dualrm{\eqalign{q^{{L(L-1)\over 2}}B^{(\nu,-1)}_{r',s'}(L,q^{-1})&=
\sum_{j=-\infty}^{\infty}(-1)^j q^{-\nu j^2-s' j}\cr
&\sum_{i=-r'}^{r'}(-1)^{r'+i}
q^{(2\nu j +s'+i)(2\nu j +s'+i-1)\over 2}
{L,2\nu j +s'+i-1;q\choose 2\nu j +s'+i}_2.}}

3) in $R^+$ as
\eqn\dualrp{\eqalign{&q^{{L(L+1)\over 2}}B^{(\nu,1)}_{r',s'}(L,q^{-1})=
{1\over 2}\sum_{j=-\infty}^{\infty}
(-1)^j q^{-\nu j^2-s' j}\cr
&\biggl(q^{(2\nu j+s'-r'-1)(2\nu j+s'-r')\over 2}
\bigl[{L,2\nu j +s' -r';q\choose 2 \nu j +s'-r'-1}_2+q^{2\nu j+s'-r'}
{L,2\nu j+s'-r'+1;q\choose 2\nu j +s'-r'}_2\bigr]\cr
&+q^{(2 \nu j +s' +r')(2\nu j +s' +r'+1)\over 2}
\bigl[{L,2 \nu j+s'+r'+1;q\choose
2\nu j +s'+r'}_2\cr
&~~~~~~~~~~~~~~~~~~~~~~~~~~~~~~~~~~~~~~~~~~~~~~+q^{2\nu j+s'+r'+1}
{L,2\nu j +s'+r'+2;q\choose 2\nu j+s'+r'+1}_2\bigr]\biggr)\cr}}

In this form we may now let $L\rightarrow \infty$ by using two limiting
results of~\rab
\eqn\dulimo{\lim _{L\rightarrow\infty}{L,A;q\choose A}_2={1\over(q)_{\infty}}}
\eqn\dulimb{\lim _{L\rightarrow\infty}{L,A-1;q\choose A}_2=
{1+q^A\over(q)_{\infty}}}
and the asymptotic formula which may be derived from~\dulimo~and (2.23) of~\rab
\eqn\dulimc{\lim_{L\rightarrow \infty}\biggl[{L,A+1;q\choose A}_2+
q^{A+1}{L,A+2;q\choose A+1}_2\biggr]={1\over (q)_{\infty}}}
to obtain for $n=0,\pm 1$
\eqn\dualres{\eqalign{(1+\theta(n>0))\lim_{L\rightarrow\infty}&
q^{L(L+n)\over 2}B^{(\nu,n)}_{r',s'}(L,q^{-1})\cr
=&q^{(s'-r')(s'-r'-|n|)\over 2}
\chi^{(2\nu-1,4\nu)}_{\nu-r'-1,2\nu-2s'-1+|n|}(q)\cr
+&q^{(s'+r'+1)(s'+r'+1-|n|)\over 2}
\chi_{\nu+r',2\nu-2s'-1+|n|}^{(2\nu-1,4\nu)}(q).\cr}}

The equation~\dualres~demonstrates that in the limit $L\rightarrow\infty$
the model $SM(2,4\nu)$ is related to  the model $M(2\nu-1,4\nu)$
by the dual transformation $q\rightarrow {1\over q}.$ This latter nonunitary
minimal model is a special case of the models $M(p,p')$ studied in ~\rbm.
It is of interest to note that while dual polynomials~\dualns--\dualrp~
yield $M(2\nu-1,4\nu)$ characters in the limit  $L\rightarrow\infty$, the
dual polynomials themselves are not the same as those of~\rfb~and ~\rbm. This
emphasizes the fact that there are many different polynomial expressions
which yield the same character in the $L\rightarrow\infty$ limit.
Indeed the polynomials of this paper and those of ~\rbm~ must be
different because in ~\rbm~ the $M(2\nu-1,4\nu)$ polynomials transform
into the $M(2\nu+1,4\nu)$ polynomials while the polynomials
{}~\dualns--\dualrp~transform into $SM(2,4\nu)$ ones.

Curiously enough, the $SM(2,8)$ model is, in fact, self-dual
\eqn\twodualns{\lim_{L\rightarrow\infty}q^{{L^2\over 2}}
B^{(2,0)}_{0,s'}(L,q^{-1})=
q^{{s'\over 2}}\bigl(\chi_{1,3-2s'}^{(3,8)}(q)+
q^{{1\over 2}+s'}\chi_{2,3-2s'}^{(3,8)}(q)\bigr)=
q^{{s'\over 2}}{\hat\chi}_{1,3-2s'}^{(2,8)}(q)}

\eqn\twodualr{\eqalign{\lim_{L\rightarrow \infty}q^{{L(L\pm 1)\over 2}}
B^{(2,\pm 1)}_{0,s'}(L,q^{-1})&=
{2\over 3\pm 1}q^{{s'-1\over 2}}\bigl(\chi_{1,4-2s'}^{(3,8)}(q)+
q\chi_{2,4-2s'}^{(3,8)}(q)\bigr)\cr
&={2\over 3\pm 1}q^{{s'-1\over 2}}(1+q\delta_{s',0})
{\hat\chi}_{1,2+2s'}^{(2,8)}(q)\cr}}
where $s'=0,1$.

To complete the study of $q$-duality we transform  the fermionic sums
using the relation
\eqn\bindual{{n+m\atopwithdelims[] m}_{q^{-1}}=
q^{-mn}{n+m\atopwithdelims[] m}_q.}
We then obtain fermionic sums with a quadratic form matrix of the type
discussed in ~\rbm. In particular, we consider $\nu\times\nu$ matrix
${\bf B}$ defined by its matrix elements
\eqn\bmat{({\bf B})_{j,k}=\cases{2&for $j=k=1$\cr
\delta_{k,2}&for $j=1,~2\leq k\leq\nu$\cr
\delta_{j,2}&for $k=1,~2\leq j\leq\nu$\cr
{1\over 2}\delta_{j,2}\delta_{k,2}+\delta_{j,k}-{1\over 2}\delta_{j,k+1}-
{1\over 2}\delta_{j,k-1}-{1\over 2}\delta_
{j,\nu}\delta_{k,\nu}&otherwise.}}
We also define
\eqn\mtilde{{\tilde{\bf m}}^t=(n_1,m_2,m_3,\cdots ,m_{\nu})}
and the $\nu-1$-dimensional vector ${\vec v}^{(k)}$
\eqn\vdef{({\vec v}^{(k)})_i=k\theta(1\leq i\leq\nu-k-1)+
(\nu-1-i)\theta(k>0)\theta(\nu-k-1<i\leq\nu-1)}
where $k=0,1,\cdots,\nu-1$.
Then, the $q$-duality transform of the fermionic polynomials
{}~\fpoly~can be expressed as
\eqn\dualfermi{\eqalign{q^{{L(L+n)\over 2}}F_{r',s'}^{(\nu,n)}(L,q^{-1})&=
\sum_{{\cal D}_{r',s'}}
q^{\Phi_n({\tilde{\bf m}},r',s')}
\prod_{i=1}^{\nu}{n_i+m_i\atopwithdelims[]n_i}_q\cr}}
where $n$, ${\cal D}_{r',s'}$ are given by~\ndefn,~\extra~and
$\Phi_n({\tilde{\bf m}},r',s')$ is defined as
\eqn\kunya{\Phi_n({\tilde{\bf m}},r',s')={1\over 2}{\tilde{\bf m}}{\bf B}
{\tilde{\bf m}}+L_n({\tilde{\bf m}},s')+C_n(r',s')}
\eqn\dulf{2L_n({\tilde{\bf m}},s')={\tilde m}_{\nu}-{\tilde m}_{\nu-s'}
+{\tilde m}_1\delta_{s',\nu-1}+(2{\tilde m}_1+{\tilde m}_2)
(n+\delta_{s',\nu-1})}
\eqn\duct{4C_n(r',s')=s'-r'+(1+2n)\delta_{s',\nu-1}.}

We now let $L\rightarrow\infty$ to obtain the following
\eqn\limdfermix{\eqalign{\lim_{L\rightarrow\infty}
q^{{L(L+n)\over 2}}F_{r',s'}^{(\nu,n)}(L,q^{-1})&=
\sum_{k=0}^{{\rm min}[r',s']}
\sum_{{\tilde{\bf m}}{\rm -restrictions}[k]}
{q^{\Phi_n({\tilde{\bf m}},r',s')}\over
(q)_{{\tilde m}_1}(q)_{{\tilde m}_2}}\cr
&\times \biggl\{\delta_{k,\nu-2}+\theta(\nu-3\geq k)
\prod_{i=3}^{\nu}{((1-{\bf B}){\tilde{\bf m}})_i-a_i^{(r')}-a_i^{(s')}
\atopwithdelims[]{\tilde m}_i}_{c,q}\biggr\}\cr}}
where the inhomogeneous vectors $a_i^{(s')}$ and $a_i^{(r')}$ are given by
{}~\indefn~ the restrictions$[k]$ on the summation variables
${\tilde{\bf m}}$ are
\eqn\restrik{{\tilde m}_i-{\tilde m}_{\nu}=
\bigl({\vec v}^{(s')}+{\vec v}^{(r')}
\bigr)_{i-1}({\rm mod}~2);~~~~~ i=2,3,\cdots,\nu-k-1;~~k\neq\nu-2}
\eqn\restro{{\tilde m}_{\nu-j}=-2(k-j);~~~~~j=0,1,2,\cdots,k\neq 0.}
and the symbol ${A\atopwithdelims[]B}_{c,q}$ in~\limdfermix~stands for the
conventional $q$-binomial coefficient (i.e. it vanishes if either $A$ or $B$
takes on negative values). Remarkably, it turns out that the formula
{}~\limdfermix~can be simplified as
\eqn\limdfermi{\eqalign{\lim_{L\rightarrow\infty}
q^{{L(L+n)\over 2}}F_{r',s'}^{(\nu,n)}(L,q^{-1})&=
\sum_{{\tilde{\bf m}}{\rm -restrictions}[0]}
{q^{\Phi_n({\tilde{\bf m}},r',s')}\over
(q)_{{\tilde m}_1}(q)_{{\tilde m}_2}}\cr
&\times \prod_{i=3}^{\nu}{((1-{\bf B}){\tilde{\bf m}})_i-a_i^{(r')}-a_i^{(s')}
\atopwithdelims[]{\tilde m}_i}_q.\cr}}

Combining results~\dualfermi,~\limdfermi~and~\polyidenta--\polyidentc~one
derives the following Fermi/Bose $M(2\nu-1,4\nu)$ character identities
\eqn\eqna{\eqalign{\lim_{L\rightarrow\infty}q^{{L^2\over 2}}
F_{r',s'}^{(\nu,0)}(L,q^{-1})&=
q^{(s'-r')^2\over 2}\chi^{(2\nu-1,4\nu)}_{\nu-r'-1,2\nu-2s'-1}(q)\cr
&+q^{(s'+r'+1)^2\over 2}\chi^{(2\nu-1,4\nu)}_{\nu+r',2\nu-2s'-1}(q)\cr}}

\eqn\eqnb{\eqalign{\lim_{L\rightarrow \infty}q^{{L(L-1)\over 2}}
F_{r',\nu-1}^{(\nu,-1)}(L,q^{-1})&=
q^{(\nu-r'-1)(\nu-r'-2)\over 2}\chi^{(2\nu-1,4\nu)}_{\nu-r'-1,2}(q)\cr
&+q^{(\nu+r')(\nu+r'-1)\over 2}\chi^{(2\nu-1,4\nu)}_{\nu+r',2}(q)\cr}}

\eqn\eqnc{\eqalign{\lim_{L\rightarrow \infty}q^{{L(L-1)\over 2}}
F_{r',s'}^{(\nu,-1)}(L,q^{-1})&=
q^{(s'-r')(s'-r'-1)\over 2}\chi^{(2\nu-1,4\nu)}_{\nu-r'-1,2\nu-2s'}(q)\cr
&+q^{(s'+r')(s'+r'+1)\over 2}\chi^{(2\nu-1,4\nu)}_{\nu+r',2\nu-2s'}(q)\cr
&+q^{(s'-r'+1)(s'-r')\over 2}\chi^{(2\nu-1,4\nu)}_{\nu-r'-1,2\nu-2s'-2}(q)\cr
&+q^{(s'+r'+1)(s'+r'+2)\over
2}\chi^{(2\nu-1,4\nu)}_{\nu+r',2\nu-2s'-2}(q),
{}~~~~~s'\neq\nu-1.\cr}}

\eqn\eqnd{\eqalign{\lim_{L\rightarrow \infty}q^{{L(L+1)\over2}}
&\biggl(F_{r',s'}^{(\nu,1)}(L,q^{-1})+
F_{r',s'-1+\delta_{s',0}}^{(\nu,1)}(L,q^{-1})\biggr)\cr
&=
q^{(s'+r')(s'+r'+1)\over 2}\chi^{(2\nu-1,4\nu)}_{\nu+r',2\nu-2s'}(q)+
q^{(s'-r')(s'-r'-1)\over 2}\chi^{(2\nu-1,4\nu)}_{\nu-r'-1,2\nu-2s'}(q).\cr}}

We would like to point out that the eqns.~\eqna~and~\eqnb~are consistent
with the results obtained in ~\rbm~ whenever $r'$ or $s'$ is equal to $0$. In
the general case identities~\eqna~and~\eqnb~are new. They demonstrate how two
quantum groups describing braiding properties of the conformal blocks
"interact" on the character level. Identities~\eqnc~and~\eqnd~are also new.
It is of interest to ascertain whether or not these new identities can be
obtained by means of Bailey Lattice technique~\dmbres,~\rfqb.

We conclude this section with the following observation. It appears that there
exist $RG$ flows connecting dual regimes of the same model. In particular, it
was proposed in~\rfatzam~that dual regimes $Z_{\nu-1}$ and $M(\nu,\nu+1)$
of the $ABF$ model~\rabf~are $RG$ connected as
\eqn\rgflow{Z_{\nu-1}+\psi_1{\bar\psi}_1+\psi^{\dag}_1{\bar\psi}^{\dag}_1
\longrightarrow M(\nu,\nu+1).}
Recently, the duality $M(p,p')\Longleftrightarrow M(p'-p,p')$ established
in~\rfb,~\rbm~was given the following $RG$ interpretation in~\rtat (see
also~\rrst)
\eqn\rgcon{M(p,p')+\phi_{2,1}\rightarrow M(p'-p,p')}
It is thus plausible that one can find an appropriate operator which would
generate a $RG$ flow connecting $SM(2,4\nu)$ and $M(2\nu-1,4\nu)$.

\newsec{On the Combinatorial bases}

It is well known that each side of a Rogers-Ramanujan type identity can be
interpreted as a generating function for a certain set of restricted partitions
{}~\rande. In particular, let $B_{\nu,s'}(N)$ denote the number of partitions
of
$N$ into parts $\neq 2({\rm mod}~ 4)$ and $\neq 0,\pm(2\nu-1-2s')
({\rm mod}~ 4\nu)$,
and $F_{\nu,s'}(N)$ denote the number of partitions of $N$ of the form
\eqn\form{N=\sum_{i=1}^{\infty}if_i}
where
\eqn\ffi{f_1+f_2\leq\nu-s'-1,~~~~~f_{2i-1}\leq 1,~~~~~
f_{2i}+f_{2i+1}+f_{2i+2}\leq\nu-1.}
Then $F_{s'}^{(\nu,0)}(q^2)\bigl(B_{s'}^{(\nu,0)}(q^2)\bigr)$ is a generating
function for $F_{\nu,s'}(N)\bigl(B_{\nu,s'}(N)\bigr)$.
Moreover, according to~\rand,~\rbres, eqn.~\bfns~implies
\eqn\threex{F_{\nu,s'}(N)=B_{\nu,s'}(N).}
By analogy with the analysis given in~\rfeino, Melzer~\rmel~proposed a
representation theoretical interpretation of~\threex~which we rephrase
as follows.

\noindent
Let $\mid{\hat\Delta}^{(2,4\nu)}_{1,2\nu-2s'-1}>$ be the highest-weight
state of  conformal dimension ${\hat\Delta}^{(2,4\nu)}_{1,2\nu-2s'-1}$
in the Verma module of $NS$ sector of $SM(2,4\nu)$. Then the set of states
\eqn\basis{W_m^{f_m}\cdots W_2^{f_2}W_1^{f_1}
\mid{\hat\Delta}^{(2,4\nu)}_{1,2\nu-2s'-1}>}
form a basis for the irreducible highest-weight representation. Here
\eqn\wdef{W_i=\cases{L_{-i\over 2},~~~~~i\equiv {\rm even} \cr
G_{-i\over 2},~~~~~i\equiv {\rm odd}.\cr}}
$L_i,G_i$ are the standard generators of the $N=1$ super-Virasoro algebra and
$f_i$ are the same as in~\ffi.

Motivated by the partition identities due to Burge (theorems 1 and 2 in
{}~\rburge) we would like to propose a different basis construction for
$SM(2,4\nu)$.

Let us introduce a set of states
\eqn\anbasis{G_{-{m\over 2}}^{f_m}\cdots G_{-{2\over 2}}^{f_2}
G_{-{1\over 2}}^{f_1}\mid{\hat\Delta}^{(2,4\nu)}_{1,2\nu-2s'-1+n}>}
where $n=0(1)$ corresponds to the $NS(R)$ sector and $f_i$ are defined as
\eqn\bffi{f_1\leq2(\nu-s'-n),~~~~~f_{2i-n}\equiv 0({\rm mod} 2),
{}~~~~~f_i+f_{i+1}\leq 2(\nu-1)}
with $s'\neq 0$ for $n=1$.

\noindent
We conjecture that the above set forms an irreducible basis. Note that our
proposal includes both $NS$ and $R$-sectors.

\noindent
To prove the above conjectures, it is sufficient to show that sets~\basis~and
{}~\anbasis~are linearly independent. Thus, one is led to the important open
question of finding an analog of the Kac determinant for
the restricted partitions.

\newsec{Concluding remarks}

We expect that it would be straightforward to apply the techniques and methods
developed here to study general $N=1$ $SM(p,p')$ character identities.
Extension to
other $SU(2)$ cosets would call for higher spin analogs of $q$-trinomial
coefficients whose properties are at present poorly understood. It is
important to find
partition theoretical and configuration sum interpretations
of~\polyidenta--\polyidentc. We believe that this interpretation would provide
important clues leading to Boltzman weights for new integrable models which
would have $SM(2,4\nu)$ and $M(2\nu-1,4\nu)$ as dual regimes. It would also be
interesting to find a $q$-trinomial generalization of Bailey's lemma.
We hope to address these challenging questions in our future publications.

\bigskip

\noindent
{\bf Acknowledgments.}
We are pleased to thank G. Andrews, W. Eholzer, A. Honecker, V. Rittenberg,
M. R{\"o}sgen, and O. Warnaar for many stimulating discussions.
We are also grateful to G. Andrews for an advance copy of~\andf.
One of us (BMM) wishes to thank Prof. J. Maillard for his hospitality at
Laboratoire de Physique Th\'eorique et des Hautes Energies located at
Universit\'e
de Paris 6-7 where part of this work was done. This work was partially
supported by NSF grant DMR 9404747 and by Deutsche Forschungs Gemeinschaft.

\appendix{A}{Proofs of Identities for $q$-trinomial Coefficients}

In this appendix we prove the identities~\identa,~\sam,~\identh~of
$q$-trinomial coefficients and the limiting formulas~\tzlim,~\dva.
We follow the notation of~\rab--\randc.

The proof of~\identa~is practically the same as that of (2.3) of~\randc.
We first use~\tndf~to rewrite~\identa~as
\eqn\appd{\eqalign{{L,A-n;q\atopwithdelims() A}_2&
=q^{L-A}{L-1,A-1-n;q\atopwithdelims()
A-1}_2+q^{L+A-n}{m-1,A-n+1;q\atopwithdelims() A+1}_2\cr
&+{L-1,A-n;q\atopwithdelims()
A}_2+q^{L-1-n}(1-q^{L-1}){L-2,A-n;q\atopwithdelims() A}_2.\cr}}
Then, recalling the definitions~\appa--\appaa~and~\qden~one can easily derive
\eqn\appf{q^{L-A}t_n(L-1,A-1;j)={q^{L-A-j}-q^L\over 1-q^L}t_n(L,A;j)}
\eqn\appg{q^{L+A-n}t_n(L-1,A+1;j-1)={q^{L-j}-q^L\over 1-q^L}t_n(L,A;j)}
\eqn\apph{t_n(L-1,A;j)={1-q^{L-2j-A}\over 1-q^L}t_n(L,A;j)}
\eqn\appi{q^{L-n-1}(1-q^{L-1})t_n(L-2,A;j-1)={q^{L-2j-A}\over 1-q^L}
(1-q^j)(1-q^{(j+A)})t_n(L,A;j).}
Then it is clear that
\eqn\appj{\eqalign{&q^{L-A}t_n(L-1,A-1;j)+q^{L+A-n}t_n(L-1.A+1;j-1)\cr
&~~~+t_n(L-1,A;j)+q^{L-n-1}(1-q^{L-1})t_n(L-2,A;j-1)\cr
&=\biggl({q^{L-A-j}-q^L\over 1-q^L}+{q^{L-j}-q^L\over 1-q^L}\cr
&~~~+{1-q^{L-2j-A}\over 1-q^L}+{q^{L-2j-A}(1-q^j)(1-q^{j+A})\over 1-q^L}
\biggr)t_n(L,A;j)\cr
&=t_n(L,A;j)\cr}}
from which~\appd~follows by summing over $j.$

We now prove~\sam. To this end we note that both sides of~\sam~satisfy the
same equation~\identa~with $n=1$. To conclude the proof one needs to verify
that~\sam~holds true for $L=0,1$. This can be easily done by the direct
inspection.

Let us now consider~\identh. This identity can be obtained from the slightly
more general identity
\eqn\appnew{\eqalign{T_{n-1}(L,A;q^{{1\over 2}})&-
T_{n-1}(L,A-2;q^{{1\over 2}})\cr
&=q^{L+A\over 2}T_{n}(L,A;q^{{1\over 2}})-q^{L+2-A\over 2}
T_n(L,A-2;q^{{1\over 2}})\cr}}
by setting $n=1$ and letting $A\rightarrow -A.$ We first use~\tndf~to rewrite
{}~\appnew~as
\eqn\appk{\eqalign{{L,A-n;q \atopwithdelims() A}_2&-
q^A{L,A-n+1;q\atopwithdelims()A}_2\cr
&=q^{1+n-A}\biggl[{L,A-2-n;q\atopwithdelims()
A-2}_2-{L,A-1-n;q\atopwithdelims() A-2}_2\biggr].}}
Then, noting the identities
\eqn\appl{t_n(L,A;j)-q^At_{n-1}(L,A;j)=\
{q^{j(j+A-n)}(q)_L\over (q)_j (q)_{j+A-1}(q)_{L-2j-A}}}
and
\eqn\appm{\eqalign{&q^{1+n-A}\biggl[t_n(L,A-2;j+1)-t_{n-1}(L,A-2;j+1)\biggr]\cr
&~~~~={q^{j(j+A-n)}(q)_L\over (q)_j (q)_{j+A-1}(q)_{L-2j-A}}\cr}}
we see that
\eqn\appn{t_n(L,A;j)-q^A t_{n-1}(L,A;j)=q^{1+n-A}\biggl[
t_n(L,A-2;j+1)-t_{n-1}(L,A-2;j+1)\biggr].}
Since
\eqn\appo{t_n(L,A-2;0)-t_{n-1}(L,A-2;0)=0,}
the desired result~\appk~follows by summing~\appn~over $j.$

To prove the limiting formula~\tzlim~we slightly extend the analysis
given in ~\rab. Let us use the elementary relation
\eqn\appn{(q^{-1})_m=(-1)^m q^{{-m(m+1)\over 2}} (q)_m}
in the definitions~\appa--\tndf~to write
\eqn\appp{\eqalign{T_n(L,A;q^{{1\over 2}})&=\sum_{l \geq 0}
{q^{2l^2-nl}(q)_L\over (q)_{{(L-A-2l)\over 2}}(q)_{{(L+A-2l)\over 2}}(q)_{2l}}~
{}~~~~~~~~~~~~~~{\rm for}~L-A~{\rm even}\cr
&=q^{{(1-n)\over 2}}\sum_{l\geq 0}{q^{2l^2+(2-n)l}(q)_L
\over (q)_{{(L-A-2l-1)\over 2}}(q)_{{(L+A-2l-1)\over 2}}(q)_{2l+1}}~{\rm for}~
L-A~{\rm odd}.\cr}}
It is now trivial to take the limit
\eqn\appq{\eqalign{\lim_{L\rightarrow \infty}T_n(L,A;q^{{1\over 2}})&
={1\over(q)_{\infty}}\sum_{j\geq 0,{\rm even}}{q^{{j(j-n)\over 2}}\over (q)_j}
{}~~{\rm for}~L-A~{\rm even}\cr
&={1\over(q)_{\infty}}\sum_{j\geq 0~{\rm odd}}{q^{{j(j-n)\over 2}}\over (q)_j}
{}~~{\rm for}~L-A~{\rm odd}\cr}}
from which, using the identity (2.20) of~\rande
\eqn\appr{\sum_{j=0}^{\infty}{t^j q^{{j(j-1)\over 2}}\over(q)_j}=
\prod_{j=0}^{\infty}(1+tq^j)}
with $t=\pm q^{{(1-n)\over 2}},$ the desired result~\tzlim~is obtained.

Finally, let us prove~\dva.
Recalling~\defofg~we can express lhs of identity~\dva~as
\eqn\zoom{g(j,q)+g(j+1,q)=\lim_{L\rightarrow\infty}G(L,j,q);~~~j\in Z,~j\geq 0}
where
\eqn\zaika{\eqalign{G(L,j,q)=\sum_{l=0}^L(-1)^l\times\biggl\{
&\bigl[T_{-1}(l,j+2)
+T_{-1}(l,j)+2T_{-1}(l,j+1)\bigr]-\cr
&\bigl[T_{-1}(l,j+1)+T_{-1}(l,j-1)+2T_{-1}(l,j)\bigr]\biggr\}.}}
Taking~\sam~into account, we find
\eqn\kozel{G(L,j,q)=(-1)^{L+1}\bigl[T_1(L+1,j)-T_1(L+1,j+1)\bigr]
-\delta_{j,0}.}
Combining~\zoom,~\kozel~and~\tolim~we obtain equation
\eqn\zaya{g(j,q)+g(j+1,q)=\lim_{L\rightarrow\infty}G(L,j,q)=-\delta_{j,0}}
which proves~\dva.

\vfill

\eject
\listrefs

\vfill\eject

\bye
\end